\documentclass[aoas]{imsart}

\RequirePackage{amsthm,amsmath,amsfonts,amssymb}
\RequirePackage[authoryear]{natbib}
\RequirePackage[colorlinks,citecolor=blue,urlcolor=blue]{hyperref}
\RequirePackage{graphicx}
\usepackage[utf8]{inputenc}
\usepackage[USenglish]{babel}
\graphicspath{{figures/}}
\usepackage{enumerate}
\usepackage{url}
\usepackage{booktabs}
\usepackage{float}

\startlocaldefs
\DeclareMathOperator*{\argmin}{argmin}
\DeclareMathOperator*{\argmax}{argmax}
\DeclareMathSymbol{\shortminus}{\mathbin}{AMSa}{"39}

\endlocaldefs

\begin{document}

\begin{frontmatter}
\title{Bayesian Cox Regression for Large-scale Inference with Applications to Electronic Health Records}
\runtitle{Bayesian Cox Regression}

\begin{aug}
\author[A,B]{\fnms{Alexander Wolfgang} \snm{Jung}\ead[label=e1,mark]{alexwjung@ebi.ac.uk}}
\and
\author[A,C,D]{\fnms{Moritz} \snm{Gerstung}\ead[label=e2,mark]{moritz.gerstung@ebi.ac.uk}}

\address[A]{European Bioinformatics Institute, EMBL-EBI \printead{e1,e2}}
\address[B]{University of Cambridge}
\address[C]{German Cancer Research Center, dkfz}
\address[D]{Genome Biology Unit, EMBL}

\end{aug}

\begin{abstract}
The Cox model is an indispensable tool for time-to-event analysis, particularly in biomedical research. However, medicine is undergoing a profound transformation, generating data at an unprecedented scale, which opens new frontiers to study and understand diseases. With the wealth of data collected, new challenges for statistical inference arise, as datasets are often high dimensional, exhibit an increasing number of measurements at irregularly spaced time points, and are simply too large to fit in memory. Many current implementations for time-to-event analysis are ill-suited for these problems as inference is computationally demanding and requires access to the full data at once. 
Here, we propose a Bayesian version for the counting process representation of Cox's partial likelihood for efficient inference on large-scale datasets with millions of data points and thousands of time-dependent covariates. Through the combination of stochastic variational inference and a reweighting of the log-likelihood, we obtain an approximation for the posterior distribution that factorizes over subsamples of the data, enabling the analysis in big data settings.
Crucially, the method produces viable uncertainty estimates for large-scale and high-dimensional datasets. 
We show the utility of our method through a simulation study and an application to myocardial infarction in the UK Biobank, where we characterize the multivariate effects of risk factors and replicate results from individual studies. 
Our framework extends the Cox model to new data sources like biobanks and EHR, the combination of which can provide new insights into our understanding of diseases.
\end{abstract}


\end{frontmatter}


\section{Introduction}
\label{subsec:intro}
A fundamental aspect of epidemiology is to identify and study the factors driving disease. One early example is the Framingham Heart Study, which shaped our understanding of cardiovascular outcomes and established risk factors like blood pressure and cholesterol that have become routine in standard health assessments \citep{dawber_epidemiological_1951}.

The possibilities and availability for this type of analysis have dramatically increased with the profound transformation of the biomedical sector in the last decades through digitization and the unprecedented scale of data generation. 
Electronic health records (EHR), containing the medical history of individuals systematically collected for millions of people and sometimes whole populations, provide new opportunities on how we can approach diseases.

Some recent studies utilizing large-scale EHR with around 1-20 million individuals include: \citet{clift_living_2020} and \citet{williamson_factors_2020} studying the risk factors associated with severe outcomes for Covid19, \citet{hippisley-cox_development_2017} predicting risk of cardiovascular disease or \citet{hippisley-cox_predicting_2021} estimating the risk of prostate cancer in asymptomatic men. 

The statistical method underlying these studies is the Cox partial likelihood, introduced by \citet{cox_regression_1972, cox_partial_1975}, as the main objective of interest is the time passed until the occurrence of an event.

While estimation with a simple Cox model is possible in large-scale settings, as the aforementioned examples show, it becomes infeasible for more complex analyses often encountered in EHR e.g. high-dimensional data, time-varying covariates, and multiple events. In many cases, one is interested in the evolving status of patients, particularly when multiple measurements over time are available, highlighting the need for a version of the Cox model that scales to the vast amounts of data while simultaneously being able to handle many of the statistical challenges arising in EHR data analysis.

\subsection{Related Work}
\label{subsec:rel_lit}

A comprehensive treatment of Bayesian methods for time-to-event analysis
can be found in \citet{ibrahim_bayesian_2001}, with applications in \citet{alvares_bayesian_2021}. 
Most implementations for the Cox model explicitly estimate the baseline hazard through a non-parametric prior, with choices including the Gamma process \citep{kalbfleisch_non-parametric_1978, sinha_semiparametric_1993}, Beta Process, \citep{hjort_nonparametric_1990, laud_bayesian_1998}, or correlated prior processes \citep{qiou_multivariate_1999}, but also spline functions \citep{sharef_bayesian_2010}. 
While specifying the baseline hazard has advantages, like absolute risk prediction, the traditional Cox model circumvents the explicit estimation.
\citet{hjort_nonparametric_1990} and \citet{sinha_bayesian_2003} provide justifications for a Bayesian version of the Cox model without the need to estimate the baseline hazard, closely resembling the traditional Cox model.
Recent Bayesian methods for variable selection in high-dimensional data include \citet{shin_scalable_2018} and \citet{nikooienejad_bayesian_2020}, however, they have mostly focused on fixed covariates. Generally, none of the Bayesian methods have focused on large-scale applications.

From a frequentist point of view, in the context of high-dimensional data, the $\textrm{Lasso}$ or $\ell_1$ regularization proposed by \citet{tibshirani_regression_1996, tibshirani_lasso_1997} has been a popular choice. An overview of time-to-event analysis with high-dimensional data more generally can be found in \citet{witten_survival_2010}. Computationally efficient implementations for the Cox model under penalization have been developed by \citet{friedman_regularization_2010}, \citet{simon_regularization_2011}, \citet{mittal_high-dimensional_2014} and \citet{yang_cocktail_2013}. Similar to the Bayesian versions, these methods have mainly focused on fixed covariates, while the R software package \textit{R-glmnet} by \citet{friedman_regularization_2010} and \citet{simon_regularization_2011} has recently added support for a Cox model based on the counting process representation in combination with the $\textrm{Lasso}$, enabling the inclusion of time-varying covariates. 
However, the proposed methods require the full data for optimization, limiting the application to moderately sized datasets.

More recently, approaches that factorize the likelihood over subsamples of the data have been proposed, facilitating time-to-event analysis at large scales. 
\cite{li_fast_2020} proposes a solution based on the batch screening iterative $\textrm{lasso}$ algorithm as an extension to the traditional $\textrm{Lasso}$. 
\citet{kvamme_time--event_2019} suggest a modified version of the Cox model based on a hypothetical weighting of the likelihood to justify training on subsamples via stochastic gradient descent (SGD).
\citet{tarkhan_bigsurvsgd_2020} provide a more general treatment of the combination of Cox's partial likelihood with SGD. 

While these methods can handle large-scale datasets, they cannot incorporate time-varying covariates measured at different time points, nor do they provide standard errors or other measures of uncertainty. The last method does propose a plug-in estimator or the bootstrap for standard errors, however, it is not clear how these will scale with large-scale and high-dimensional datasets. 

A divide-and-conquer approach to the Cox model has been proposed in \citet{wang2021fast} that circumvents some of the aforementioned limitations. However, the algorithm requires $n>>p$ and is generally limited to a moderate number of covariates. Given the extent of information available in EHR, the need for a method that can not only scale with the number of individuals but also explore and utilize high-dimensional covariate sets is paramount. 

\subsection{Main Contributions}
\label{subsec:con}

In this study, we propose a Bayesian Cox model for big data settings based on the flexible counting representation introduced in \citet{andersen_coxs_1982}.
The advantages of the counting process representation are described in \citet{therneau_cox_2000} and enable to fit models with time-varying covariates, complex censoring and truncation patterns, multiple time-scales, multiple events, and marginal or conditional models for correlated data. 
By combining our model with a reweighting of the log-likelihood and stochastic variational inference (SVI), we can decouple the sample size from our estimation procedure, enabling inference via subsampling of the full data and hence analyzing data that cannot fit into memory at once e.g. EHR. 
By using a sparsity inducing prior, we can also do inference in high-dimensional data.
Our approach enables the inclusion of time-varying covariates measured at irregularly spaced time points, as well as viable uncertainty estimates for large-scale and high-dimensional datasets.

\noindent The main contributions of this paper are:
\begin{itemize}
  \item Large-scale implementation of a Bayesian version of Cox's partial likelihood based on a counting process representation. 
  \item Reweighting of Cox's partial log-likelihood for subsampled data to enable stochastic variational inference 
  \item Viable uncertainty estimates for large-scale and high-dimensional datasets. 
  \item Simulation algorithm for discrete event times with time-varying covariates resembling EHR. 
  \item Application of the proposed method to myocardial infarction in the UK Biobank. This comprises a Cox regression for $\approx$400,000 individuals with $\approx$1.5 million observations and $\approx$1,000 time-varying covariates.
\end{itemize}

\noindent The rest of the paper is structured as follows:\\
In Section \ref{sec:meth} we describe the method in detail. A comprehensive simulation study resembling EHR data with a particular focus on time-varying covariates is provided in Section \ref{sec:sim}. A modified algorithm to efficiently simulate large-scale and high-dimensional event time data is described in Section \ref{subsec:simbin}. The results of the simulation are given in Section \ref{subsec:standard} and \ref{subsec:highdim}, where we compare our method with other Cox model implementations. Then, the method is evaluated on real-world applications in Section \ref{sec:app}. Section \ref{subsec:exambles} shows the model performance on a standard dataset provided in the \textit{R-Survival} package. Section \ref{subsec:ukb} is a small case study on the risk of myocardial infarction in the UK-Biobank. Section \ref{sec:lim} discusses some limitations and future directions while Section \ref{sec:dis} concludes the paper.

\section{Methods}
\label{sec:meth}

A typical analysis with EHR comprises the emergence of a disease in the population, and one wants to understand the effect of other comorbidities and potential biomarkers on it. The main interest is the time passed until the onset of a particular event, in this case a disease, denoted here as the event times $T_i$ for an individual $i$.  

Generally, one wants to understand how covariates may increase or decrease the risk of the event, consequently accelerating or decelerating the event times. The covariates are allowed to vary over time e.g. the evolving set of 
diseases an individual acquires or updated laboratory measurements, denoted as a column vector $\mathbf{X}_{i\cdot}(t) = (X_{i1}(t), \dots, X_{i\mathrm{p}}(t))^{\top}$ of the $\mathrm{p}$-covariate processes. 

Typically, we have incomplete data, like censoring, truncation, or filtering. Here, we assume for simplicity independent right censoring, e.g. physical time limit of observations in EHR. The proposed method, however, can readily be used in conjunction with more complex missing data patterns.
Formally, we have an individual censoring time $u_i$ such that we only observe ${y_i = \min\{T_i, u_i\}}$.
The observed data for $n$ individuals is then $\mathbf{D} = \{(y_i, \delta_i, \mathbf{X}_{i\cdot}(t)): i = 1,2,\dots, n\}$ with $\delta_i=1_{[T_i=y_i]}$ as an indicator for censoring.

Some important concepts from time-to-event analysis, assuming continuous time, are the survival function  ${S(t) = \mathcal{P}(T > t)}$, i.e. the probability that the event has not happened by $t$, and the hazard function ${{\alpha(t) = \lim_{\Delta t \to 0} (\Delta t)^{-1} \mathcal{P}(T \leq t + \Delta t \hspace{0.05cm} | \hspace{0.05cm} T > t)} = f(t)S(t)^{-1}}$, i.e. the instantaneous risk of observing an event given the event has not happened yet, where $f(t)$ is the density function. The cumulative hazard is given as ${A(t) = \int_0^t \alpha(u)\textrm{d}u}$, and is linked to the survival function through ${S(t) = exp(-A(t))}$.

\subsection{Cox's partial likelihood based on counting processes}
\label{subsec:counting}

For the observational model we define a multivariate counting process for the $n$ individuals in the EHR as $\mathbf{N}(t) = (N_1(t), N_2(t), \dots, N_n(t))^{\top}$, with individual counting process $N_i(t)$, counting the occurrences of events for individual $i$ with jump sizes $\Delta \mathbf{N} = 1$, assuming that no two processes jump simultaneously at $t$ (no ties).
In principle $N_i(t)$ can count multiple events e.g. recurrent cancer diagnoses, however, we treat it here as if only a single event occurs, without loss of generality. 
The intensity process, the rate at which we expect events to happen, for $N_i(t)$ is linked to the hazard function under the Cox model by 
${\lambda_i^{\boldsymbol{\gamma}, \boldsymbol{\theta}}(t) = I_i(t)\alpha_i^{\boldsymbol{\gamma}, \boldsymbol{\theta}}(t) = I_i(t)\alpha_0^{\boldsymbol{\gamma}}(t)exp\big[\boldsymbol{\theta}^{\top}\mathbf{X}_{i\cdot}(t)\big]}$ where $I_i(t)$ is a predictable process with values in $\{0, 1\}$, describing whether individual $i$ is at risk of observing the event at $t$.
The baseline hazard $\alpha_0^{\boldsymbol{\gamma}}(t)$, defined by the possibly infinite-dimensional parameter $\boldsymbol{\gamma}$, determines the rate at which events occur without any covariate effects. 
The $\mathrm{p}$-regression parameter vector $\boldsymbol{\theta}$ describes the effect of the time-varying covariates, as they act multiplicatively on the baseline hazard.

The likelihood $\mathcal{L}$ for independent right-censored event times is the combined effect of the density function for individuals observing events and the survival function for censored individuals, which can na\"ivly be written by means of the hazard and survival function as
\begin{equation*}
\mathcal{L} = \prod_i^n f_i^{\delta_i} S_i^{1 - \delta_i} = \prod_i^n \alpha_i^{\delta_i} S_i\,.
\end{equation*}.

The likelihood for $\boldsymbol{\gamma}$ and $\boldsymbol{\theta}$ under the counting process representation of the Cox model follows as
\begin{equation}
\begin{aligned}
\label{eq:likelihood}
\mathcal{L}(\mathbf{D} | \boldsymbol{\theta}, \boldsymbol{\gamma})= & \prod_{t \in \mathfrak{T}} \bigg\{ \prod_{i}^{n} \big(\alpha_0^{\boldsymbol{\gamma}}(t)\exp\big[\boldsymbol{\theta}^{\top}\mathbf{X}_{i\cdot}(t)\big]\big)^{\Delta N_i(t)}\bigg\} \\ &  exp\bigg[-\int_0^t \sum_{k=1}^{n}I_k(u) \exp\big[\boldsymbol{\theta}^{\top}\mathbf{X}_{k\cdot}(u)\big] \textrm{d}A_0^{\boldsymbol{\gamma}}(u) \bigg]\,,
\end{aligned}
\end{equation}
where $\mathfrak{T}$ is the set of event times and $A_0^{\boldsymbol{\gamma}}(t)$ is the cumulative baseline hazard. 
As we are mainly interested in $\boldsymbol{\theta}$, it is possible to derive a profile likelihood in the sense of
\begin{equation}
\label{eq:partial-likelihood}
\mathcal{L}(\mathbf{D} | \boldsymbol{\theta}) = \max_{\boldsymbol{\gamma}}\mathcal{L}(\mathbf{D} | \boldsymbol{\theta}, \boldsymbol{\gamma}) = \prod_{t \in \mathfrak{T}} \prod_i^{n} \bigg( \frac{\exp\big[\boldsymbol{\theta}^{\top}\mathbf{X}_{i\cdot}(t)\big]}{\sum_{k=1}^{n}I_k(t) \exp\big[\boldsymbol{\theta}^{\top}\mathbf{X}_{k\cdot}(t)\big]}  \bigg)^{\Delta N_i(t)}\,,
\end{equation}
which is equivalent to the partial likelihood specified by \citet{cox_regression_1972, cox_partial_1975}. For details on how to derive the profile/partial likelihood see \citet{andersen_coxs_1982}.
The log-likelihood follows as
\begin{equation}
\label{eq:loglikelihood}
\log \mathcal{L}(\mathbf{D} | \boldsymbol{\theta}) = \sum_i^n \int_0^t \boldsymbol{\theta}^{\top}\mathbf{X}_{i\cdot}(u)\textrm{d}N_i(u) - \int_0^t \log\bigg( \sum_{k=1}^{n}I_k(u) \exp\big[\boldsymbol{\theta}^{\top}\mathbf{X}_{k\cdot}(u)\big]\bigg) \textrm{d}\mathbf{N}(u) \,.
\end{equation}
For a more detailed derivation of Cox's partial likelihood under a counting process representation we refer to \citet{andersen_statistical_1993}.

\subsection{Reweighting of Cox's partial likelihood}
\label{subsec:reweighted}

As can be seen in Equation \eqref{eq:partial-likelihood} the partial likelihood cannot be factorized over the data as the denominator contains a sum over the full data. Hence, to use subsamples $\mathbf{D}^* \ll \mathbf{D}$ for inference in large-scale datasets we need to reweight the likelihood, or more precisely the log-likelihood from Equation \eqref{eq:loglikelihood} appropriately.
We reweight the log-likelihood as it will be the main objective for inference. 
To achieve a good approximation we propose the following adjustments to the partial log-likelihood \eqref{eq:loglikelihood}:
\begin{equation}
\label{eq:loglikelihood_reweigthed}
\begin{aligned}
\log \mathcal{L}(\mathbf{D} | \boldsymbol{\theta}) \approx \, & w_1 \sum_i^{n^*} \int_0^t \boldsymbol{\theta}^{\top}\mathbf{X}_{i\cdot}(u)\textrm{d}N_i(u) - \\ &  w_1 \int_0^t \log\bigg( w_2\sum_{k=1}^{n^*}I_k(u) \exp\big[\boldsymbol{\theta}^{\top}\mathbf{X}_{k\cdot}(u)\big]\bigg) \textrm{d}\mathbf{N}^*(u) \,,
\end{aligned}
\end{equation}
where $n^* \ll n$ and $w_1 = \int_0^t
\boldsymbol{1}\textrm{d}\mathbf{N}(u)(\int_0^t \boldsymbol{1}\textrm{d}\mathbf{N}^*(u))^{-1}$ i.e. the ratio of events
, $w_2 = n (n^*)^{-1}$ i.e. the ratio of observations, with $\mathbf{N}^*$ representing the multivariate counting process corresponding to the subsampled individuals $n^*$. The motivation for the reweighting is given in Section 1 of the supplementary materials.
Particularly, the last approximation i.e. the denominator of the partial likelihood  
may introduce a bias. However, we show through simulations in the supplementary materials Section 1 that the possible bias is very small for a variety of simulation and optimization settings.

\subsection{Bayesian Cox Model}
\label{sec:bayes}

To use the partial likelihood defined in Equation \eqref{eq:partial-likelihood} for probabilistic inference further justification is needed as it is not necessarily a likelihood in the classical sense. \\
\citet{kalbfleisch_marginal_1973} have shown that the partial likelihood can be interpreted as proportional to the marginal distribution of the rankings of the event times, however, the derivation does not include time-varying covariates. \citet{kalbfleisch_non-parametric_1978} provides a Bayesian justification based on an underlying diffuse Gamma process for $A_0^{\gamma}$, showing that the partial likelihood is the limiting marginal posterior distribution. This approach is further extended by \citet{sinha_bayesian_2003}, encompassing a large class of models, including (external) time-varying covariates. We show through simulations in Section \ref{sec:sim} that this provides valid estimates even for discretized internal time-varying covariate processes. \citet{hjort_nonparametric_1990} provides additional justifications for the partial likelihood based on Beta and Dirichlet processes. 

We can  define the posterior distribution following \citet{hjort_nonparametric_1990} and \citet{sinha_bayesian_2003} and assuming prior independence between $p(\boldsymbol{\theta})$ and  $p(\boldsymbol{\gamma})$ as: 
\begin{align}
\label{eq:posterior}
\begin{split}
p(\boldsymbol{\theta} | \mathbf{D}) &\propto \int \mathcal{L}(\mathbf{D} | \boldsymbol{\theta}, \boldsymbol{\gamma}) p(\boldsymbol{\theta}) p(\boldsymbol{\gamma}) \textrm{d}\boldsymbol{\gamma}\\
&\propto \mathcal{L}(\mathbf{D} | \boldsymbol{\theta}) p(\boldsymbol{\theta})\,,
\end{split}
\end{align}
where $\mathcal{L}(\mathbf{D} | \boldsymbol{\theta}, \boldsymbol{\gamma})$ is taken from Equation \eqref{eq:likelihood} and $\mathcal{L}(\mathbf{D} | \boldsymbol{\theta})$ from Equation \eqref{eq:partial-likelihood}.

As prior for $p(\boldsymbol{\theta})$ we use either the Normal distribution for a moderate number of covariates
\begin{equation*}
\theta_l \hspace{0.05cm}|\hspace{0.05cm} \sigma \sim \mathrm{Normal}(0, \sigma^2), \hspace{0.5cm} l = 1, 2, \dots, \mathrm{p}
\end{equation*}
or in the case of many covariates and high-dimensional data, the $\mathrm{Student} {\shortminus} t_{\nu}$ distribution with $\nu$ degrees of freedom to induce sparsity
\begin{equation*}
\theta_l \hspace{0.05cm}|\hspace{0.05cm} \nu, s \sim \mathrm{Student} {\shortminus} t_\nu(0, s^2), \hspace{0.5cm} l = 1, 2, \dots, \mathrm{p} \,.  
\end{equation*}

\subsection{Variational Inference}
\label{sec:variational}

While it is possible to design an efficient accept/reject algorithm to do MCMC, we recast the inference problem as optimization. Precisely, we define a family $\mathfrak{G}$ of variational distributions and try to find the member $q(\boldsymbol{\theta})$ of the family that closely approximates the posterior distribution from Equation \eqref{eq:posterior}. Closeness is measured as the Kullback-Leibler ($\mathcal{KL}$-) divergence
\begin{equation*}
\label{eq:KL}
q^*(\boldsymbol{\theta}) = \argmin_{q{(\boldsymbol{\theta})} \in \mathfrak{G}} \mathcal{KL}(q(\boldsymbol{\theta}) \hspace{0.05cm} || \hspace{0.05cm} p(\boldsymbol{\theta} | \mathbf{D})) = 
\argmin_{q{(\boldsymbol{\theta})} \in \mathfrak{G}} \hspace{0.1cm} \int q(\boldsymbol{\theta})\log\frac{q(\boldsymbol{\theta})}{p(\boldsymbol{\theta} | \mathbf{D})}\textrm{d}\boldsymbol{\theta} \,.
\end{equation*}
Generally, it is not possible to optimize the  $\mathcal{KL}$-divergence directly, instead we maximize the evidence lower bound ($\textrm{ELBO}$)
\begin{equation*}
\label{eq:ELBO}
\textrm{ELBO}(q) = \mathbb{E}_q[\log \mathcal{L}(\mathbf{D} | \boldsymbol{\theta})] + \mathbb{E}_q[\log p(\boldsymbol{\theta})] - \mathbb{E}_q[\log q(\boldsymbol{\theta})] \,,
\end{equation*}
where $\log \mathcal{L}(\mathbf{D} | \boldsymbol{\theta})$ is our reweighted version from Equation \eqref{eq:loglikelihood_reweigthed}, $p(\boldsymbol{\theta})$ the prior density and $q(\boldsymbol{\theta})$ the variational distribution. 
Our final objective for inference consequently becomes 
\begin{equation*}
\label{eq:ELBO_objective}
q^*(\boldsymbol{\theta}) = \argmax_{q{(\boldsymbol{\theta})} \in \mathfrak{G}} \textrm{ELBO}(q) \,.
\end{equation*}
An explicit formulation of the objective function and an outline of the variational inference procedure can be seen in Section 2 of the supplementary materials.

\citet{andersen_coxs_1982} have shown that the estimator of $\boldsymbol{\theta}$ under the Cox model is asymptotically Normal, hence, we propose a Multivariate Normal family for $\mathfrak{G}$ when the number of covariates is moderate, which should give a reasonable approximation, especially for large-scale settings. 

For a large number of covariates and high-dimensional data we use a lower rank approximation of the covariance matrix in the form $\boldsymbol{\Sigma} = \mathbf{W}^T\mathbf{W} + \mathbf{I}$, where $\mathbf{I}$ denotes identity matrix, and  $\mathbf{W}\in \mathbb{R}^{R\times P}$, with $R$ as the rank and $P$ the number of covariates. 
The low-rank approximation for high-dimensional data is experimental. However, we show through simulations in Section \ref{subsec:highdim} that this provides a good approximation.
Various methods have been proposed to handle difficult variational objectives by replacing the expectation in the $\textrm{ELBO}$ through Monte Carlo (MC) samples and optimization via automatic differentiation. For details on some algorithms see \citet{ranganath_black_2014} and \citet{kucukelbir_automatic_2017}. A detailed survey of MC-based gradient estimation can be found in \citet{mohamed_monte_2019}.
We use automatic differentiation variational inference (ADVI) following \citet{kucukelbir_automatic_2017} for the optimization of the $\textrm{ELBO}$, with a random subsample $\mathbf{D}^*$ at each optimization step.

\section{Simulation}
\label{sec:sim}
The overall setting for the simulations is the occurrence of a disease in the general population and the association of time-varying covariates with it, emulating a typical scenario encountered when analyzing EHR. Event times are generated based on the Cox proportional hazards model and are measured as age in days, ranging from birth to 80 years. Covariates vary randomly over time at discrete time steps and are drawn from either a Normal or Bernoulli distribution, resembling laboratory measurements or disease status. The baseline hazard reflects typical age-sex adjusted distributions of disease with an initial increase and a decrease in older ages.
Individuals are followed from birth until either censoring or an event occurs. Censoring is independent of the data generating process. 

\subsection{Simulation of event times via the Binomial model}
\label{subsec:simbin}
Our proposed algorithm for simulation of event-times follows along the lines of \citet{sylvestre_comparison_2008} and is based on the discretized partial likelihood defined by \citet{cox_regression_1972} as
\begin{equation*}
\frac{\alpha(t, \mathbf{X})\textrm{d}t}{1- \alpha(t, \mathbf{X})\textrm{d}t} = \exp\big[\boldsymbol{\theta}^{\top}\mathbf{X}(t)\big]\frac{\alpha_0(t)\textrm{d}t}{1- \alpha_0(t)\textrm{d}t} \,.
\end{equation*}
Based on this, one can define a Binomial model with the probability of an event at time $t$ as
\begin{equation}
\label{eq:discretePL}
\mathcal{P}(t=T) = (1 + \exp\big[ (\log(\alpha_0(t)) -\log(1-\alpha_0(t))) + \boldsymbol{\theta}^{\top}\mathbf{X}(t)        \big])^{-1}\,.
\end{equation} 

\begin{figure}[h!]
    \centering
    \includegraphics[width=0.90\textwidth]{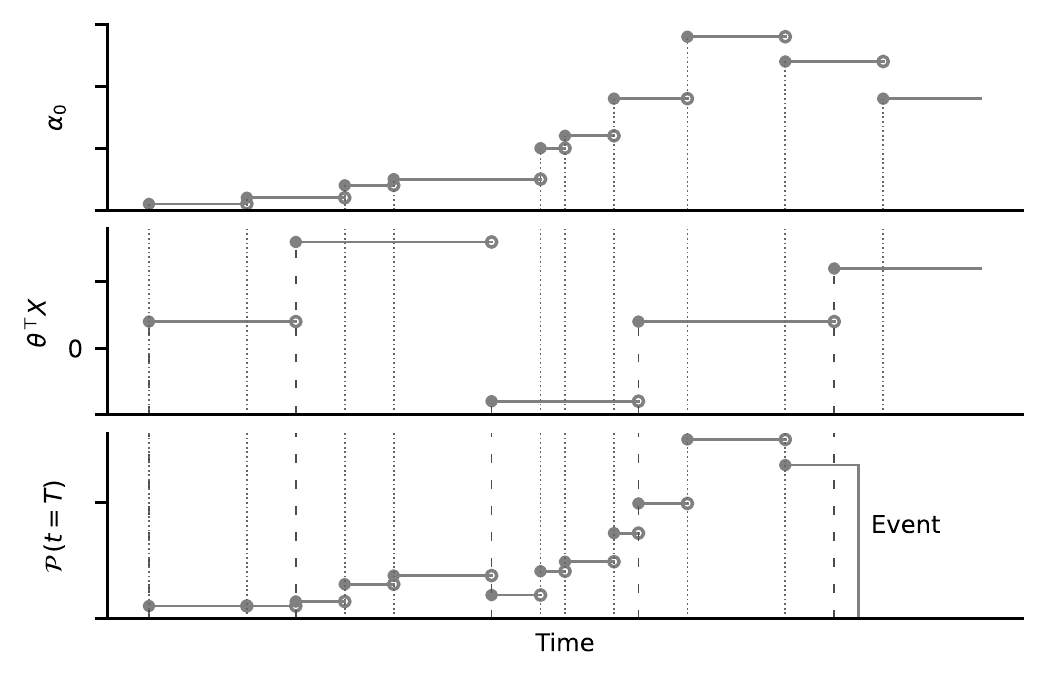}
    \caption{A schematic overview for discrete failure time simulation.
    The upper panel shows the rough shape of the baseline hazard $\alpha_0$. The middle panel represents for an individual the corresponding linear effect of the time-varying covariate processes $\mathbf{X}_{i\cdot}(t)$ and the parameters $\boldsymbol{\theta}$. The lower panel depicts the combined effect of the baseline hazard and the covariate effects. The dotted vertical line represents jump times of the baseline hazard and the dashed vertical line the jump times for the covariate processes. The probability of an event $\mathcal{P}(t=T)$ is the combination of both processes with jumps at either of the two and is evaluated for each $\textrm{d}t$ until either censoring or an event happens as indicated by the solid vertical line in the lower panel.}
    \label{fig:schematic}
\end{figure}

For a given simulation, we draw a sample of the underlying baseline hazard. Per individual, we simulate the $\mathrm{p}$-covariate processes and then move successively through time $t = 1, 2, \dots, c$, where $c$ is the individual censoring time $c \sim  \mathrm{Uniform}(1000, 30000)$.
For a given $t$ we evaluate the individual probability of an event as in Equation \eqref{eq:discretePL} conditional on the state of the baseline hazard and the covariate processes at $t$ until either an event occurs or the censoring time is reached. The covariate processes up to the final time are then retrieved and kept for the subsequent analysis.
A schematic representation of the simulation can be seen in Figure \ref{fig:schematic}.

For the simulations presented in this study, the covariate processes and the baseline hazard resemble a sample path from a renewal-reward process $\{Z(t, \omega), t=0, 1, \dots\}$ with $\omega$ as a particular realization, given the waiting times $W_1, W_2, \dots$ between the jumps of the process $J_{k} = \sum_{i=1}^{k}W_i$ and $Z(t) = \sum_{i=1}^{\infty} R_i \cdot \, \mathbb{I}_{J_i \leq t} $ with $\mathbb{I}_{J_i \leq t}$ as the indicator function and $R_i$ as the rewards.
Here, the time between increments $W_i$ are rounded to the nearest integer.

For the covariate processes $X_{ip}(t) = Z(\cdot, \omega)$ with $W_1, W_2, \dots \sim \mathrm{Gamma}(4, 500)$ and $R_i \sim \mathrm{Normal}(0, 1)$ for continuous and $R_i \sim \mathrm{Bernoulli}(0.2)$ for binary variables. 

The baseline hazard $\alpha_0(t) = Z(\cdot, \omega)$ with $W_1, W_2, \dots \sim \mathrm{Gamma}(4, 200)$ and 
\begin{equation*}
R_i \sim 
\begin{cases}
    +\mathrm{Gamma}(2, 1),  & \text{if } J_i \leq 15000 \,, \\
    +\mathrm{Gamma}(1, 10), & \text{if } 15000 < J_i \leq 25000 \,, \\
    -\mathrm{Gamma}(1, 5),  & \text{otherwise.}
\end{cases}
\end{equation*}

The sampled baseline hazard can then be scaled up or down to reflect the appropriate rate of censorship. The resulting shape for the baseline hazard is similar to the depiction in the schematic. 
An example output for two individuals in the resulting long-format can be seen in Table \ref{tab:example}.

\begin{table}[h!]
\caption{Example output for a simulation draw of 2 individuals and 11 observations with 3 binary and 3 quantitative covariates in long format.}
\label{tab:example}
\centering
\resizebox{0.60\textwidth}{!}{%
\begin{tabular}{@{}ccccccccc@{}}
\hline\hline \addlinespace[0.75ex]
start & stop  & event & X1 & X2 & X3 & X4    & X5    & X6    \\ \midrule
0     & 1000  & 0     & 0  & 0  & 0  & 0     & 0     & 0     \\
1000  & 3235  & 0     & 1  & 0  & 0  & -1.22 & -0.3  & 0.61  \\
3235  & 6671  & 0     & 0  & 0  & 1  & 0.14  & 0.32  & 0.61  \\
6671  & 8551  & 0     & 0  & 0  & 0  & -0.76 & -0.2  & 0.65  \\
8551  & 10146 & 0     & 0  & 1  & 1  & -0.19 & -0.52 & 0.13  \\
10146 & 11107 & 1     & 0  & 1  & 0  & 1.55  & 0.85  & 1.84  \\
0     & 3693  & 0     & 0  & 0  & 0  & 0     & 0     & 0     \\
3693  & 5421  & 0     & 0  & 0  & 0  & -0.42 & -1.87 & 0.34  \\
5421  & 7291  & 0     & 0  & 0  & 0  & 0.64  & -2.43 & 0.51  \\
7291  & 9852  & 0     & 0  & 0  & 0  & 0.09  & 0.03  & 0.71  \\
9852  & 10004 & 0     & 0  & 0  & 0  & -1.23 & 0.76  & -0.03 \\ 
\addlinespace[0.75ex]
\hline\hline
\end{tabular}%
}
\end{table}

\subsection{Log-likelihood approximation through reweighting}
\label{subsec:llapprox}
First, we wanted to evaluate the subsampled approximation of the full log-likelihood through the reweighting described in Equation \eqref{eq:loglikelihood_reweigthed}, and particularly, how it compares to a more na\"ive approximation that would simply scale $\log \mathcal{L}$ by the relative number of observations between the full data and the batch size. 
The data has been generated as:

\underline{Simulation:}
\noindent $2000\,\,individuals, \;\;15883\,\,observations, \;\;0.90\,\, censorship_{ind},
\newline 5000\,\,covariates\;\;(2500\,\,binary,\;\;2500\,\,continuous), \;\; \theta_l \sim \mathrm{Normal}(0, 0.5^2), \;\;  \newline 1024\,\,
Batch\,size_{obs}, \;\; \hat{\theta_l} \sim \mathrm{Student} {\shortminus} t(\nu=1, s=0.001),\;\; Rank=50$.

\begin{figure}[h!]
    \centering
    \includegraphics[width=0.95\textwidth]{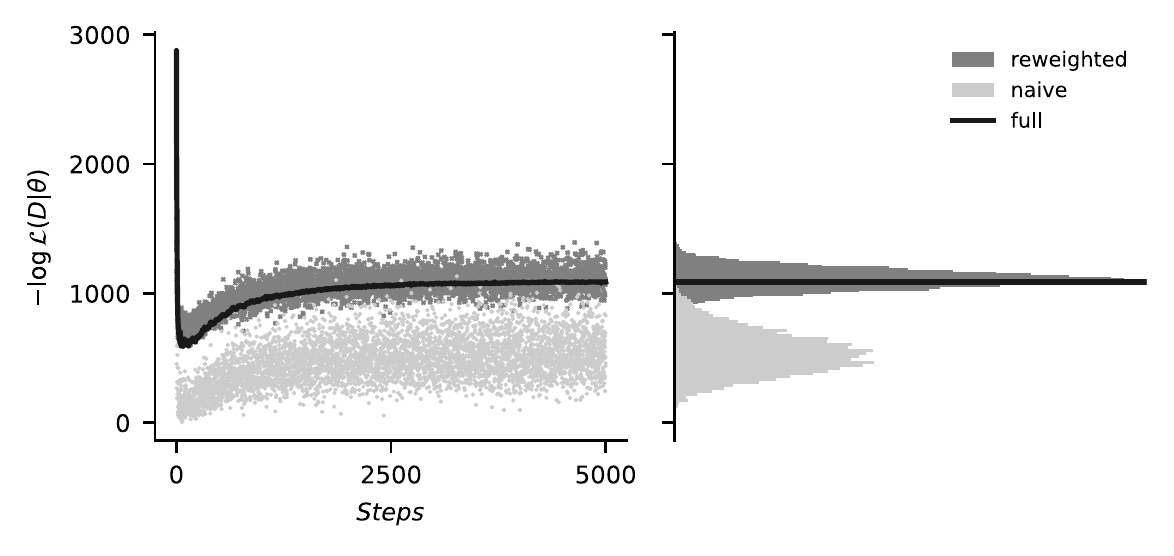}
    \caption{Log-likelihood approximation along training. On the left-hand side, we can see the full log-likelihood evaluated at each training step in black, the corresponding reweighted log-likelihood evaluations in dark grey and a na\"ive approximation in light grey. The right-hand side shows the distributions of the approximations around the full-likelihood evaluated at the last training step for repeated samples.}
    \label{fig:gradflow_highdim}
\end{figure}

The results can be seen in Figure \ref{fig:gradflow_highdim}.
The reweighting forms an almost unbiased, symmetric distribution around the full log-likelihood throughout the training process and is substantially better then a na\"ive approximation based on the batch size itself. Interesting to note here is the initial drop in the -log-likelihood followed by an increase - showing the effect of penalization. A similar depiction for a standard case can be seen in Section 1 of the supplementary materials.
Additionally, we evaluate the performance of our reweighting for different scenarios based on different levels of observations, batch size, censorship, number of covariates, and hazard. Overall the approximation of the log-likelihood is very good with almost no error on average. 

\subsection{Standard case}
\label{subsec:standard}
The standard case simulation reflects the typical data structure most often encountered in much of applied medical research with a small number of time-varying covariates and a moderately sized population such that $observations$ $\gg$ $individuals$ $\gg$ $covariates$. This setting allows us to assess individual parameter estimates based on bias, efficiency, and coverage. Overall we run two separate simulations for different parameter values and varying censorship. The specific settings for the simulations are: 

\underline{Simulation 1}: $200\,\,runs, \;\;1000\,\,individuals, \;\;7213-7906\,\,observations, \newline 0.74-0.83\,\,censorship_{ind}, \;\;0.0-0.02\,\,ties, \;\;6\,\,covariates\;\;(3\,\,binary,\;\;3\,\,continuous), \newline
\boldsymbol{\theta} = (-0.9, 0.2, 0.0, -0.4, 1.1, 0.0)^{\top}.$

\underline{Simulation 2}: $200\,\,runs, \;\;2000\,\,individuals, \;\;15605-16844\,\,observations, \newline 0.92-0.96\,\,censorship_{ind}, \;\;0.0-0.03\,\,ties, \;\;6\,\,covariates\;\;(3\,\,binary,\;\;3\,\,continuous), \newline
\boldsymbol{\theta} = (0.8, -0.5, 0.0, -0.7, 1.0, 0.0)^{\top}.$

\begin{table}[h!]
\caption{Simulation results for the standard case setting. The upper panel shows the results for the first simulation and the lower panel represents the second simulation. The first 3 estimates correspond to the binary covariates, the last 3 to the quantitative covariates.}

\label{tab:standard}
\centering
\resizebox{0.99\textwidth}{!}{%
\begin{tabular}{@{}ccccccclccccc@{}}
\hline\hline
\addlinespace[0.75ex]
                     &                      & \multicolumn{5}{c}{ProbCox - $Batch\,size_{obs} 256$}                                                                                     &  & \multicolumn{5}{c}{R-Survival}                                                                                                          \\ \cmidrule[\heavyrulewidth]{2-7} \cmidrule[\heavyrulewidth]{9-13}
$\theta$             &                      & ${\hat{\theta}}$ & ${\sigma_{\hat{\theta}}}$ & $RMSE$               & $\overline{HPD}_{95\%}$ & $Coverage_{95\%}$    &  & ${\hat{\theta}}$ & ${\sigma_{\hat{\theta}}}$ & $RMSE$               & $\overline{CI}_{95\%}$ & $Coverage_{95\%}$    \\ \cmidrule{2-7} \cmidrule{9-13}
-0.9                 &                      & -0.89                & 0.22                               & 0.22                 & 0.98                    & 0.94                 &  & -0.92                & 0.23                               & 0.23                 & 0.91                   & 0.94                 \\
0.2                  &                      & 0.18                 & 0.17                               & 0.17                 & 0.74                    & 0.98                 &  & 0.2                  & 0.16                               & 0.16                 & 0.64                   & 0.97                 \\
0                    &                      & -0.04                & 0.18                               & 0.18                 & 0.78                    & 0.96                 &  & -0.02                & 0.18                               & 0.18                 & 0.68                   & 0.94                 \\
-0.4                 &                      & -0.39                & 0.07                               & 0.07                 & 0.32                    & 0.96                 &  & -0.39                & 0.06                               & 0.07                 & 0.27                   & 0.96                 \\
1.1                  &                      & 1.09                 & 0.08                               & 0.08                 & 0.34                    & 0.96                 &  & 1.1                  & 0.07                               & 0.07                 & 0.28                   & 0.96                 \\
0                    &                      & 0.01                 & 0.07                               & 0.07                 & 0.31                    & 0.98                 &  & 0.01                 & 0.07                               & 0.07                 & 0.27                   & 0.94                 \\
\multicolumn{1}{l}{} & \multicolumn{1}{l}{} & \multicolumn{1}{l}{} & \multicolumn{1}{l}{}               & \multicolumn{1}{l}{} & \multicolumn{1}{l}{}    & \multicolumn{1}{l}{} &  & \multicolumn{1}{l}{} & \multicolumn{1}{l}{}               & \multicolumn{1}{l}{} & \multicolumn{1}{l}{}   & \multicolumn{1}{l}{} \\
0.8                  &                      & 0.74                 & 0.2                                & 0.21                 & 0.85                    & 0.96                 &  & 0.79                 & 0.19                               & 0.19                 & 0.72                   & 0.93                 \\
-0.5                 &                      & -0.52                & 0.25                               & 0.25                 & 1.11                    & 0.96                 &  & -0.49                & 0.24                               & 0.24                 & 1.01                   & 0.98                 \\
0                    &                      & -0.05                & 0.23                               & 0.23                 & 0.99                    & 0.96                 &  & -0.01                & 0.23                               & 0.23                 & 0.87                   & 0.94                 \\
-0.7                 &                      & -0.69                & 0.11                               & 0.11                 & 0.41                    & 0.92                 &  & -0.69                & 0.1                                & 0.1                  & 0.35                   & 0.9                  \\
1                    &                      & 0.99                 & 0.09                               & 0.09                 & 0.43                    & 0.96                 &  & 0.99                 & 0.08                               & 0.08                 & 0.35                   & 0.97                 \\
0                    &                      & 0                    & 0.09                               & 0.09                 & 0.4                     & 0.94                 &  & 0                    & 0.09                               & 0.09                 & 0.34                   & 0.97                 \\ 
\addlinespace[0.75ex]
\hline\hline
\addlinespace[0.75ex]
\multicolumn{13}{l}{\begin{tabular}[c]{@{}l@{}}Notes: RMSE - Root mean squared error, HPD - Highest posterior density, CI - Confidence interval\end{tabular}}
\end{tabular}
}
\end{table}

We use a Multivariate Normal family for variational inference and $\hat{\theta_l} \sim \mathrm{Normal}(0, 1)$ as prior. 
The results for the simulation are shown in Table \ref{tab:standard}. 
We provide results for the Cox model fitted with the implementation \textit{R-Survival} in \citet{therneau_package_2021} on the full data and our implementation \textit{ProbCox} on subsamples. The results for different batch sizes can be found in the supplementary materials Section 3. 
The average parameter estimates $\hat{\theta}$ are consistent across the different settings and stable for different batch sizes. The variability of the estimates across the simulations measured as the standard deviation $\sigma_{\hat{\theta}}$ as well as the root mean squared error RMSE are comparable to \textit{R-Survival}. The coverage is close to the nominal rate with the average highest posterior density interval length $\overline{HPD}_{95\%}$ similar in size to the average confidence interval length $\overline{CI}_{95\%}$ from \textit{R-Survival}, increasing with smaller batches, reflecting the additional approximation error through subsampling. Overall, we see comparable performance between our implementation trained on subsamples and \textit{R-Survival} fitted on the full data.

\subsection{High-dimensional case}
\label{subsec:highdim}
With the number of variables measured on individuals continuously increasing the need for models that can handle large numbers of covariates and high-dimensional data ($covariates$ $\gg$ $observations$ $\gg$ $individuals$) is paramount. Particularly, when analyzing EHR the subsampled data for each evaluation can effectively be very high-dimensional, even though the EHR itself is not. Hence, a method that can scale to EHR will also need to be able to handle high-dimensional data. 
Therefore, we evaluated our method on a very high-dimensional dataset ($covariates$ $\gg$ $observations$) and a moderate high-dimensional dataset ($covariates$ $\gg$ $individuals$) with strong correlation between covariates. The simulations have been conducted as follows:

\underline{Simulation 1}: $\; 200\,\,runs, \;\;1000\,\,individuals, \;\;6896-7613\,\,observations, \newline 0.65-0.72\,\,censorship_{ind}, \;\;0.0-0.02\,\,ties, \;\;
\boldsymbol{\theta}_{1:20} \sim \mathrm{Normal}(0, 0.75^2), \;\;\boldsymbol{\theta}_{21:10000} = \boldsymbol{0} \newline 
10000\,\,covariates\;\;(5000\,\,binary,\;\;5000\,\,continuous). \newline
$
Parameters are equally split between binary and continuous covariates.

\begin{figure}[H]
    \centering    \includegraphics[width=0.85\textwidth]{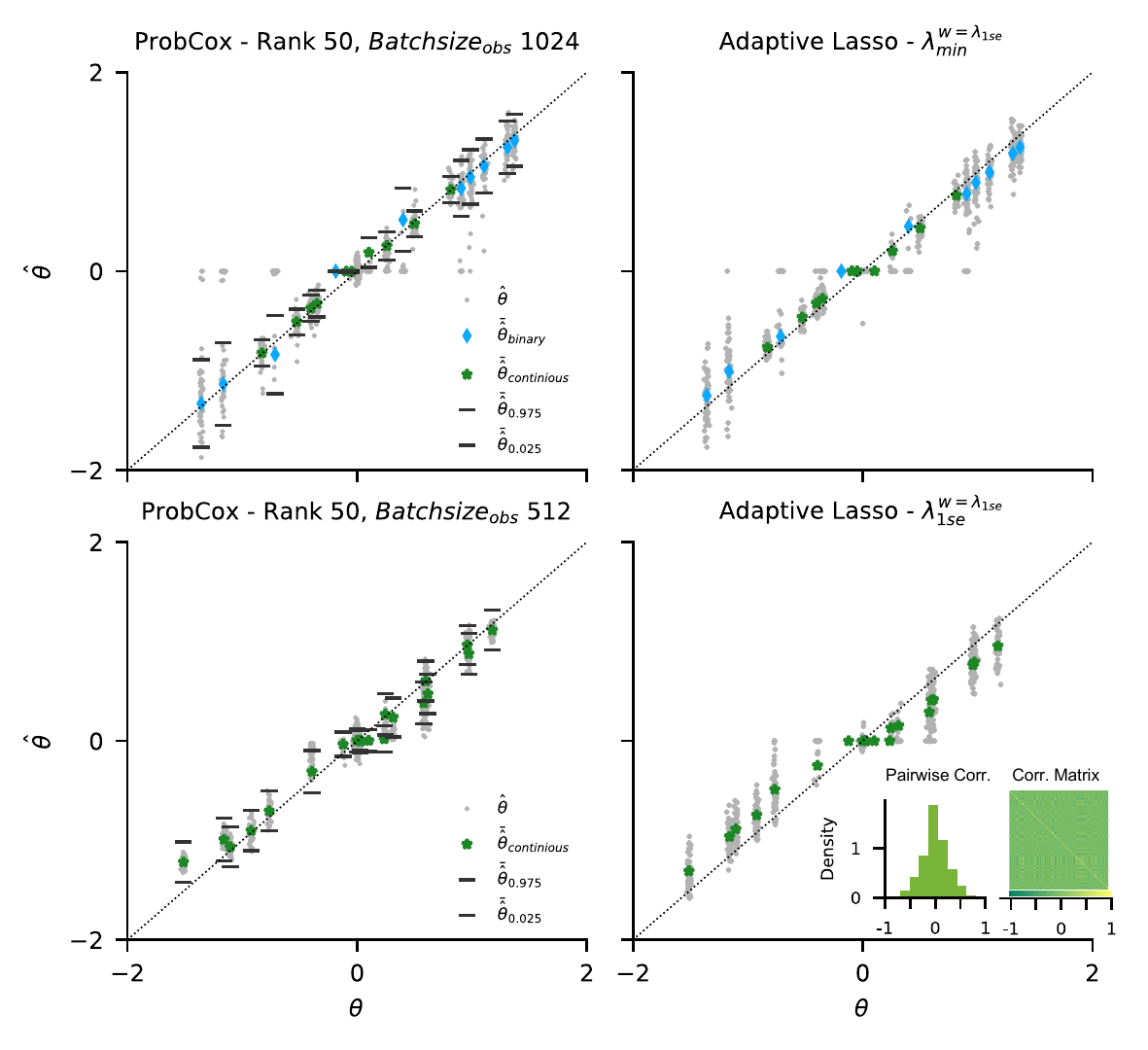}
    \caption{Simulation results for the high-dimensional case. The first panel corresponds to simulation 1 and the second panel to simulation 2. The light grey dots are parameter estimates from individual simulation runs, while the diamonds and stars represent the average estimate over all simulations for binary and continuous variables, respectively. The black bars are the average ${HPD}_{95\%}$ bounds. The dotted line corresponds to the diagonal and indicates correct identification of the true parameters}
    \label{fig:highdim}
\end{figure}

\underline{Simulation 2}: $\; 200\,\,runs, \;\;1000\,\,individuals, \;\;6803-7469\,\,observations, \newline 0.66-0.74\,\,censorship_{ind}, \;\;0.41-0.54\,\,ties, \;\;
\boldsymbol{\theta}_{1:20} \sim \mathrm{Normal}(0, 0.75^2), \;\;\boldsymbol{\theta}_{21:3000} = \boldsymbol{0} \newline 
3000\,\,covariates\;\;(0\,\,binary,\;\;3000\,\,continuous)\;\;. \newline
$
Correlations have been randomly induced following \citet{lewandowski2009generating} and the underlying correlation matrix as well as the distribution of pairwise correlations can be seen in Figure \ref{fig:highdim}.

The results for the simulations can be seen in Table \ref{tab:highdim}.
We provide a comparison of our proposed method to the Cox model with the Adaptive Lasso described in \citet{zou_adaptive_2006} and \cite{zhang_adaptive_2007}. The model is fit via the \textit{R-glmnet} package.
The estimation procedure is two-fold. First, we run an initial Cox regression with Lasso penalty where the 
$\lambda$ parameter (strength of penalization) is chosen through 5-fold cross-validation on the concordance index. The coefficients corresponding to either $\lambda_{1se}$ or $\lambda_{min}$ are extracted. Coefficients estimated to be 0 are replaced with a value of 1e-06 and subsequently used as the weights for the Adaptive Lasso (similar settings as in the first run). Typically, the weights would be estimated with a Ridge penalty, however, this did not provide good results, and therefore, we used the Lasso for the weight estimation as well.
We show results for the Adaptive Lasso with $\lambda_{min}^{w=\lambda_{1se}}$ and $\lambda_{1se}^{w=\lambda_{1se}}$ for simulation 1 and simulation 2, respectively, as they showed the best performance for parameter identification. 
For our method we use the low rank approximation to the Multivariate Normal family for variational inference and ${\hat{\theta_l} \sim \mathrm{Student} {\shortminus} t(\nu=1, s=0.001)}$ as prior. 
A parameter is na\"ivly classified as identified if the $HPD_{95\%}$ does not contain zero.
It should be noted that this is not a proper variable selection procedure and we use this just as a rough guide to compare identification of parameters between the methods. 

For the Adaptive Lasso a parameter is identified if the estimate is not equal to zero. The rate of identification for the parameters over all simulations is given in the last column of Table \ref{tab:highdim} for each method, respectively. 

Summaries are computed over the set of estimates that have been identified in the respective simulation runs. 

\begin{table}[h!]
\caption{Simulation results for the high-dimensional case - non-zero parameters only. The first panel corresponds to simulation 1 and the second panel to simulation 2. For simulation 1, the first ten rows correspond to the non-zero binary covariates and the last ten rows to the non-zero continuous covariates.}
\label{tab:highdim}
\centering
\resizebox{0.99\textwidth}{!}{%
\begin{tabular}{@{}ccccccccccccccc@{}}
\hline\hline
\addlinespace[0.75ex]
         &  & \multicolumn{6}{c}{ProbCox - $Rank\,50, \; Batch\,size_{obs}\,1024$}                                                                                   & \multicolumn{1}{l}{} & \multicolumn{6}{c}{Adaptive Lasso - $\lambda_{min}^{w=\lambda_{1se}}$}                                                                      \\ \cmidrule[\heavyrulewidth]{3-8}
\cmidrule[\heavyrulewidth]{10-15}
$\theta$ &  & $\bar{\hat{\theta}}$ & $\overline{\sigma_{\hat{\theta}}}$ & $RMSE$ & $\overline{HPD}_{95\%}$ & $Coverage_{95\%}$ & $p_{|\hat{\theta}| > 0}$ &                      & $\bar{\hat{\theta}}$ & $\overline{\sigma_{\hat{\theta}}}$ & $RMSE$ & $\overline{HPD}_{95\%}$ & $Coverage_{95\%}$ & $p_{|\hat{\theta}| > 0}$ \\
\cmidrule{3-8}
\cmidrule{10-15}
-0.71    &  & -0.78                & 0.16                               & 0.17   & 0.76                    & 0.97              & 0.29                     &                      & -0.66                & 0.23                               & 0.24   & -                       & -                 & 0.26                     \\
1.31     &  & 1.25                 & 0.14                               & 0.15   & 0.51                    & 0.92              & 1                        &                      & 1.18                 & 0.14                               & 0.18   & -                       & -                 & 1                        \\
1.37     &  & 1.33                 & 0.14                               & 0.15   & 0.5                     & 0.92              & 1                        &                      & 1.25                 & 0.13                               & 0.18   & -                       & -                 & 1                        \\
0.91     &  & 0.85                 & 0.15                               & 0.16   & 0.54                    & 0.92              & 0.92                     &                      & 0.78                 & 0.16                               & 0.21   & -                       & -                 & 0.9                      \\
0.4      &  & 0.53                 & 0.06                               & 0.14   & 0.56                    & 1                 & 0.06                     &                      & 0.45                 & 0.18                               & 0.19   & -                       & -                 & 0.06                     \\
-0.19    &  & 0                    & 0.04                               & 0.19   & 0.02                    & 0                 & 0                        &                      & 0                    & 0                                  & 0.19   & -                       & -                 & 0                        \\
0.99     &  & 0.95                 & 0.14                               & 0.14   & 0.53                    & 0.94              & 0.98                     &                      & 0.89                 & 0.16                               & 0.18   & -                       & -                 & 0.98                     \\
1.1      &  & 1.06                 & 0.15                               & 0.16   & 0.52                    & 0.92              & 1                        &                      & 0.99                 & 0.16                               & 0.19   & -                       & -                 & 0.99                     \\
-1.36    &  & -1.3                 & 0.24                               & 0.24   & 0.83                    & 0.92              & 0.95                     &                      & -1.25                & 0.27                               & 0.3    & -                       & -                 & 0.98                     \\
-1.16    &  & -1.1                 & 0.2                                & 0.21   & 0.79                    & 0.93              & 0.87                     &                      & -1.01                & 0.24                               & 0.29   & -                       & -                 & 0.96                     \\
-0.83    &  & -0.82                & 0.08                               & 0.08   & 0.25                    & 0.94              & 1                        &                      & -0.77                & 0.07                               & 0.09   & -                       & -                 & 1                        \\
0.1      &  & 0.19                 & 0.04                               & 0.1    & 0.26                    & 0.85              & 0.06                     &                      & 0                    & 0.01                               & 0.1    & -                       & -                 & 0                        \\
0.26     &  & 0.26                 & 0.07                               & 0.07   & 0.27                    & 0.97              & 0.7                      &                      & 0.2                  & 0.08                               & 0.09   & -                       & -                 & 0.32                     \\
0.82     &  & 0.82                 & 0.09                               & 0.09   & 0.25                    & 0.92              & 1                        &                      & 0.76                 & 0.07                               & 0.09   & -                       & -                 & 1                        \\
-0.53    &  & -0.51                & 0.08                               & 0.08   & 0.25                    & 0.94              & 1                        &                      & -0.46                & 0.07                               & 0.1    & -                       & -                 & 1                        \\
-0.4     &  & -0.38                & 0.08                               & 0.08   & 0.25                    & 0.9               & 1                        &                      & -0.32                & 0.08                               & 0.12   & -                       & -                 & 0.91                     \\
0.5      &  & 0.49                 & 0.08                               & 0.08   & 0.25                    & 0.94              & 1                        &                      & 0.43                 & 0.07                               & 0.1    & -                       & -                 & 1                        \\
-0.35    &  & -0.33                & 0.07                               & 0.08   & 0.26                    & 0.94              & 0.96                     &                      & -0.28                & 0.09                               & 0.11   & -                       & -                 & 0.76                     \\
-0.09    &  & -0.22                & 0.07                               & 0.14   & 0.28                    & 0.67              & 0.04                     &                      & 0                    & 0                                  & 0.09   & -                       & -                 & 0                        \\
-0.05    &  & 0                    & 0.01                               & 0.05   & 0.02                    & 0.05              & 0                        &                      & 0                    & 0                                  & 0.05   & -                       & -                 & 0                        \\ 
\addlinespace[1.75ex]
\hline
\addlinespace[1.75ex]
         &  & \multicolumn{6}{c}{ProbCox - $Rank\,50, \; Batch\,size_{obs}\,512$}                                                                                                      &  & \multicolumn{6}{c}{Adaptive Lasso - $\lambda_{1se}^{w=\lambda_{1se}}$}                                                                      \\ \cmidrule[\heavyrulewidth]{3-8}
\cmidrule[\heavyrulewidth]{10-15}
$\theta$ &  & $\bar{\hat{\theta}}$ & $\overline{\sigma_{\hat{\theta}}}$ & $RMSE$ & $\overline{HPD}_{95\%}$ & $Coverage_{95\%}$ & $p_{|\hat{\theta}| > 0}$ &  & $\bar{\hat{\theta}}$ & $\overline{\sigma_{\hat{\theta}}}$ & $RMSE$ & $\overline{HPD}_{95\%}$ & $Coverage_{95\%}$ & $p_{|\hat{\theta}| > 0}$ \\ \cmidrule[\heavyrulewidth]{3-8}
\cmidrule[\heavyrulewidth]{10-15}
0.1      &  & 0                    & 0.01                               & 0.1                & 0.22                    & 0.94              & 0                        &  & 0                    & 0                                  & 0.1                                                & -                       & -                 & 0                        \\
0.31     &  & 0.24                 & 0.06                               & 0.1                & 0.39                    & 1                 & 0.06                     &  & 0.15                 & 0.11                               & 0.19                                               & -                       & -                 & 0.22                     \\
0.97     &  & 0.88                 & 0.09                               & 0.13               & 0.41                    & 0.86              & 1                        &  & 0.79                 & 0.14                               & 0.23                                               & -                       & -                 & 1                        \\
-0.39    &  & -0.31                & 0.08                               & 0.11               & 0.42                    & 1                 & 0.3                      &  & -0.25                & 0.11                               & 0.19                                               & -                       & -                 & 0.13                     \\
-0.77    &  & -0.71                & 0.09                               & 0.11               & 0.4                     & 0.95              & 1                        &  & -0.49                & 0.19                               & 0.34                                               & -                       & -                 & 0.96                     \\
0.24     &  & 0.27                 & 0.05                               & 0.06               & 0.41                    & 1                 & 0.14                     &  & 0.13                 & 0.08                               & 0.14                                               & -                       & -                 & 0.22                     \\
0.96     &  & 0.96                 & 0.08                               & 0.08               & 0.39                    & 1                 & 1                        &  & 0.76                 & 0.13                               & 0.24                                               & -                       & -                 & 1                        \\
0.62     &  & 0.47                 & 0.12                               & 0.19               & 0.4                     & 0.7               & 0.9                      &  & 0.42                 & 0.15                               & 0.25                                               & -                       & -                 & 0.98                     \\
0.03     &  & 0.01                 & 0.02                               & 0.03               & 0.2                     & 1                 & 0                        &  & 0                    & 0                                  & 0.03                                               & -                       & -                 & 0                        \\
-0.92    &  & -0.9                 & 0.08                               & 0.08               & 0.41                    & 0.98              & 1                        &  & -0.75                & 0.13                               & 0.22                                               & -                       & -                 & 1                        \\
-1.51    &  & -1.22                & 0.04                               & 0.3                & 0.41                    & 0.02              & 1                        &  & -1.31                & 0.16                               & 0.26                                               & -                       & -                 & 1                        \\
0.03     &  & 0                    & 0.02                               & 0.04               & 0.23                    & 1                 & 0                        &  & 0                    & 0                                  & 0.03                                               & -                       & -                 & 0                        \\
-1.1     &  & -1.07                & 0.06                               & 0.08               & 0.4                     & 1                 & 1                        &  & -0.89                & 0.13                               & 0.26                                               & -                       & -                 & 1                        \\
-1.16    &  & -0.99                & 0.07                               & 0.18               & 0.43                    & 0.76              & 1                        &  & -0.96                & 0.14                               & 0.24                                               & -                       & -                 & 1                        \\
1.18     &  & 1.11                 & 0.06                               & 0.09               & 0.4                     & 0.99              & 1                        &  & 0.95                 & 0.16                               & 0.27                                               & -                       & -                 & 1                        \\
0        &  & 0                    & 0.01                               & 0.01               & 0.24                    & 1                 & 0                        &  & 0                    & 0                                  & 0                                                  & -                       & -                 & 0                        \\
0.24     &  & 0.02                 & 0.02                               & 0.22               & 0.26                    & 0.01              & 0                        &  & 0                    & 0.01                               & 0.24                                               & -                       & -                 & 0.01                     \\
0.6      &  & 0.6                  & 0.11                               & 0.11               & 0.4                     & 0.94              & 0.99                     &  & 0.41                 & 0.14                               & 0.24                                               & -                       & -                 & 0.99                     \\
-0.12    &  & -0.03                & 0.04                               & 0.1                & 0.24                    & 0.61              & 0.02                     &  & 0                    & 0.01                               & 0.12                                               & -                       & -                 & 0                        \\
0.58     &  & 0.38                 & 0.1                                & 0.22               & 0.42                    & 0.51              & 0.74                     &  & 0.29                 & 0.16                               & 0.33                                               & -                       & -                 & 0.52                    
                    \\ \hline\hline
\multicolumn{15}{l}{\begin{tabular}[c]{@{}l@{}}Notes: RMSE - Root mean squared error, HPD - Highest posterior density, $p_{|\hat{\theta}|}$ -  probability of identification - \textit{R-glmnet} estimates for the standard error are not available.  \end{tabular}}
\end{tabular}%
}
\end{table}

Generally, the estimates of our method work well in both simulation settings, being on average close to the true parameter values and exhibiting good coverage overall. The results are robust across the rank and batch size specifications.
In comparison to the Adaptive Lasso, our estimates are slightly closer to the true parameters. For identification, both methods perform equally well with both approaches identifying on average 13 covariates correctly and 1-2 miss-identified and 11 covariates correctly and 1 miss-identified for simulation 1 and simulation 2, respectively. 

A graphical comparison of our method and the Adaptive Lasso can be seen in Figure \ref{fig:highdim}.
Most of our estimates lie on the diagonal enclosed by the average ${HPD}_{95\%}$ bounds, indicating again a good identification of the true parameters on average with good quantification of the uncertainty. The estimates for adaptive Lasso are also close to the diagonal.

An overview and additional results for the Adaptive Lasso, Lasso, and our method with different batch sizes and ranks can be found in the supplementary materials Section 4. 
 
We provide an additional high-dimensional simulation comparing our method again to the Adaptive Lasso, and additionally to the smoothly clipped absolute deviation (SCAD) penalty, the minimax concave penalty (MCP) implemented in \textit{R-ncvreg} by \citet{breheny_coordinate_2011}, as well as another Bayesian approach with inverse moment priors implemented in \textit{R-BVSNLP} by \citet{nikooienejad_bayesian_2020}.
However, implementations for these do only allow for simple Cox models and consequently no time-varying covariates. The simulation and results can be found in the supplementary materials Section 5. Overall, our method performs best in parameter estimation and is comparable in identification. 

We conducted another high-dimensional simulation which can be found in Section 6 of the supplementary materials, to further evaluate and assess the performance of our method under different hyperparameter settings for the $\mathrm{Student} {\shortminus} t$ distribution as well as a comparison to the Bayesian Lasso as described in \citet{park_bayesian_2008} and \citet{hans2009bayesian}.

Also, we run a large-scale simulation to evaluate the potential bias when the ratio between batch size to observations is very low ($\approx 0.0001$). The simulation and results can be found in Section 7 of the supplementary materials. Parameter estimates are close to the true values on average with good coverage.

\subsection{Resource comparison}
\label{subsec:reso}

\begin{figure}[h!]
    \centering
    \includegraphics[width=0.6\textwidth]{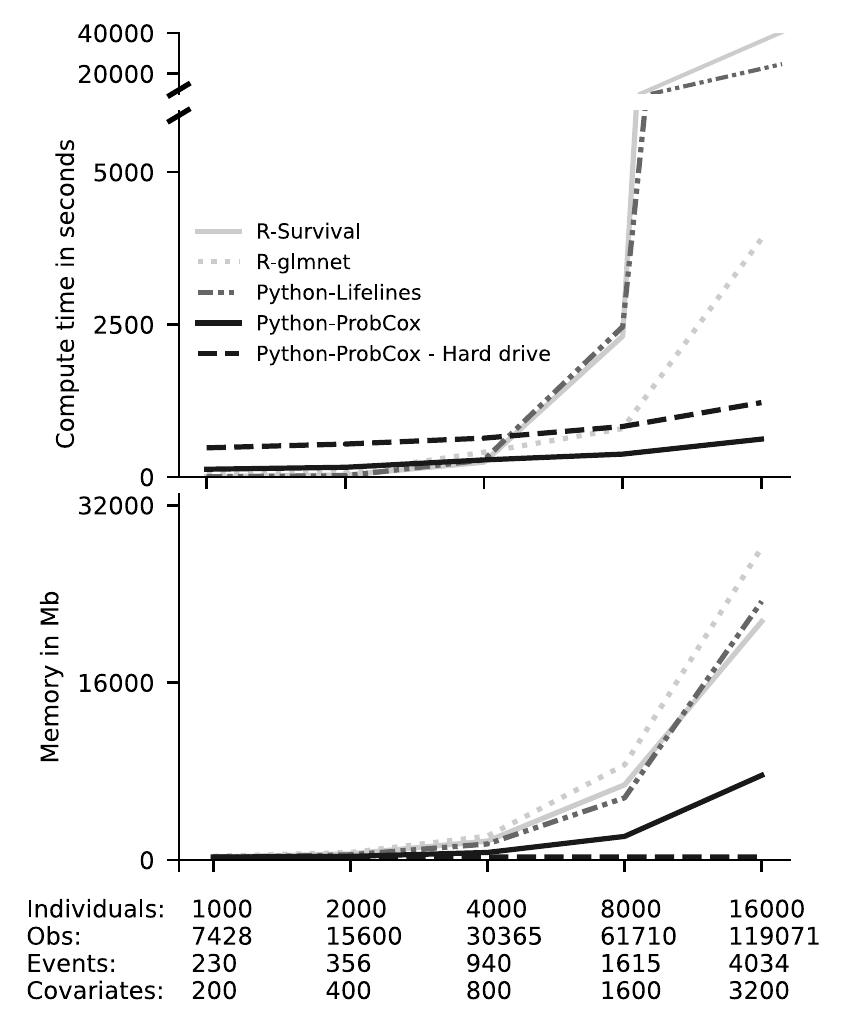}
    \caption{Runtime/Memory comparison between Python-ProbCox, Python-ProbCox (Hard drive), Python-lifelines, \textit{R-Survival}, and \textit{R-glmnet}. ProbCox (Hard drive) loads the relevant batch sample on demand from the hard drive instead of loading the full data into memory.}
    \label{fig:run}
\end{figure}

In this section, we provide a small compute-time and memory utilization comparison between our method \textit{Python-ProbCox}, \textit{R-Survival}, \textit{R-glmnet}, and \textit{Python-lifelines}.
We run two different versions of our method, one that feeds the subsamples from memory, and one version that loads each subsample from the hard drive, reflecting the more typical use-case of our method. 
Both version are run with the low rank approximation ($R=25$) to the Multivariate Normal family for variational inference, ${\hat{\theta_l} \sim \mathrm{Student} {\shortminus} t(\nu=1, s=0.001)}$ as prior, and a batch size of 256 observations.
The main characteristics for the simulation can be taken from Figure \ref{fig:run}. Covariates are equally split between binary and continuous covariates of which ten are non-zero with ${\boldsymbol{\theta}_{1:10} \sim \mathrm{Normal}(0, 0.5^2)}$. 
We only run a single run per simulation per method. Therefore, the estimates only serve as a rough guide. 
The timings and memory utilization for a 4-core machine can be seen in Figure \ref{fig:run}. 

Overall, we can see a good resource utilization of our method. There is an initial overhead which is offset once the dataset becomes large-scale and high-dimensional. We can also see a clear trade-off between memory usage and compute time. However, the additional compute time levied through loading of the subsamples from the hard-drive is neglectable and allows for almost constant memory usage. 

Additionally, we performed a separate comparison of our method to the implementation in \citet{wang2021fast}. Both methods perform well in regard to parameter identification and estimation and are close in speed for large-scale cases. The Analysis can be found in Section 8 of the supplementary materials.

\section{Application}
\label{sec:app}
\subsection{Small-scale examples} 
\label{subsec:exambles}

To further evaluate our proposal and compare it with the standard implementations, we run a couple of small applications on real-world data provided in \textit{R-Survival}. The focus is on comparing the methods rather than on the data analysis.
In total, we compare 5 different datasets, the results for 4 of them can be found in the supplementary material Section 9. Our proposed method compares very well with the parameter and standard error estimates being similar.

The \textit{PBCseq} data contains 312 individuals with 1,945 observations and 725 events where we treat repeated events as independent for simplicity. For missing covariates, we carry the information from the previous visit forward. Additionally, we construct an artificial high-dimensional dataset based on the \textit{PBCseq} data by appending 2,000 $\mathrm{Normal}(0, 1)$ distributed covariates. The continuous covariates from the \textit{PBCseq} have been Z-transformed accordingly.
We compare our method to \textit{R-Survival} for the original dataset and the Adaptive Lasso implemented through \textit{R-glmnet} for the high-dimensional dataset.
We use a Multivariate Normal family for variational inference with ${\theta_l} \sim \mathrm{Normal}(0, 1)$ as prior for the original data and the low rank ($R=50$) approximation to the Multivariate Normal family with ${{\theta_l} \sim \mathrm{Student} {\shortminus} t(\nu=1, s=0.001)}$ in the high-dimensional case. The batch size for both cases is 256 observations. The results can be see in Table \ref{tab:app}.

 \begin{table}[h!]
 \caption{Results for the \textit{PBCseq} application. The first two columns correspond to the original and the last two columns to the high-dimensional analysis.}
\label{tab:app}
\centering
\resizebox{0.99\textwidth}{!}{%
\begin{tabular}{@{}lccccccccccc@{}}
\hline\hline
\addlinespace[0.75ex]
          & \multicolumn{2}{c}{ProbCox} & \multicolumn{1}{l}{} & \multicolumn{2}{c}{R-Survival} & \multicolumn{1}{l}{} & \multicolumn{2}{c}{ProbCox - HD} & \multicolumn{1}{l}{} & \multicolumn{2}{c}{Adaptive Lasso - $\lambda_{min}^{w=\lambda_{1se}}$} \\

          \cmidrule[\heavyrulewidth]{2-3}
          \cmidrule[\heavyrulewidth]{5-6}
          \cmidrule[\heavyrulewidth]{8-9}
          \cmidrule[\heavyrulewidth]{11-12}

Var:      & $\theta$  & $HPD_{95\%}$    &                      & $\theta$    & $CI_{95\%}$      &                      & $\theta$     & $HPD_{95\%}$      &                      & $\theta$    & $CI_{95\%}$    \\
          \cmidrule{2-3}
          \cmidrule{5-6}
          \cmidrule{8-9}
          \cmidrule{11-12}
trt       & -0.07     & (-0.22, 0.08)   &                      & -0.07       & (-0.22, 0.08)    &                      & 0.           & (-)               &                      & 0.          & -              \\
sex       & 0.58*     & (0.39, 0.78)    &                      & 0.59*       & (0.39, 0.78)     &                      & 0.54*        & (0.34, 0.74)      &                      & 0.43        & -              \\
edema=0   & -0.55*    & (-0.82, -0.28)  &                      & -0.55*      & (-0.82, -0.28)   &                      & -0.29*        & (-0.48, -0.1)   &                      & -0.31       & -              \\
edema=0.5 & -0.3*     & (-0.47, -0.13)  &                      & -0.3*       & (-0.55, -0.05)   &                      & -0.          & (-)               &                      & 0.          & -              \\
stage=2   & -0.12     & (-0.59, 0.35)   &                      & -0.13       & (-0.66, 0.4)     &                      & -0.          & (-)               &                      & 0.          & -              \\
stage=3   & -0.03     & (-0.29, 0.23)   &                      & -0.07       & (-0.57, 0.43)    &                      & 0.           & (-)               &                      & 0.          & -              \\
stage=4   & 0.08      & (-0.09, 0.25)   &                      & 0.05        & (-0.45, 0.54)    &                      & -0.          & (-)               &                      & 0.          & -              \\
ascites   & -0.09     & (-0.31, 0.13)   &                      & -0.12       & (-0.36, 0.12)    &                      & 0.           & (-)               &                      & 0.          & -              \\
hepato    & 0.2*      & (0.03, 0.37)    &                      & 0.18*       & (0.0, 0.36)      &                      & 0.17         & (-0.03, 0.38)     &                      & 0.          & -              \\
spiders   & 0.08      & (-0.07, 0.24)   &                      & 0.07        & (-0.09, 0.24)    &                      & -0.          & (-)               &                      & 0.          & -              \\
bili      & 0.3*      & (0.24, 0.37)    &                      & 0.31*       & (0.24, 0.38)     &                      & 0.33*        & (0.26, 0.4)       &                      & 0.37        & -              \\
chol      & -0.05     & (-0.12, 0.02)   &                      & -0.05       & (-0.12, 0.02)    &                      & -0.          & (-)               &                      & 0.          & -              \\
albumin   & -0.25*    & (-0.33, -0.17)  &                      & -0.26*      & (-0.35, -0.17)   &                      & -0.29*       & (-0.37, -0.21)    &                      & -0.27       & -              \\
alk.phos  & 0.18*     & (0.11, 0.24)    &                      & 0.18*       & (0.11, 0.25)     &                      & 0.15*        & (0.08, 0.22)      &                      & 0.          & -              \\
platelet  & 0.03      & (-0.08, 0.14)   &                      & -0.01       & (-0.1, 0.08)     &                      & 0.           & (-)               &                      & 0.          & -              \\
protime   & 0.15*     & (0.09, 0.21)    &                      & 0.15*       & (0.09, 0.22)     &                      & 0.15*        & (0.09, 0.22)      &                      & 0.14        & -              \\
age       & 0.28*     & (0.21, 0.35)    &                      & 0.28*       & (0.2, 0.36)      &                      & 0.28*        & (0.2, 0.35)       &                      & 0.26        & -              \\
          &           &                 &                      &             &                  &                      &              &                   &                      &             &                \\
          \cmidrule{2-12}

Harrel's C:  & \multicolumn{2}{c}{0.729}   &                      & \multicolumn{2}{c}{0.729}                &                      & \multicolumn{2}{c}{0.741}                  &                      & \multicolumn{2}{c}{0.726}           \\ 
\addlinespace[0.75ex]
\hline\hline
\addlinespace[0.75ex]
\multicolumn{12}{l}{Notes: * non-zero effect at 95\%, (-) interval indistinguishable from 0 after rounding}
\end{tabular}%
}
\end{table}

For the original data, our method is again comparable to the implementation in \textit{R-Survival} with the same covariates identified as having non-zero effects and similar estimates for parameters and uncertainty. For the high dimensional case, we identify similar covariates as the Adaptive Lasso with them being the significant covariates from the original analysis. The estimates are close to the parameter estimates from the original evaluations for both approaches. Overall our method identified 7 non-zero covariates (based on the $HPD_{95\%}$) and the Adaptive Lasso 6 non-zero covariates.

\subsection{Myocardial infarction in the UK Biobank}
\label{subsec:ukb}
In this section, we showcase a typical example of analysis where we see the greatest utility of our proposed method. With the number of biobanks and population-based health registers continuously increasing we suspect these types of analyses to become more widespread and relevant. Being able to conduct multivariate association studies at such a large scale provides new possibilities to identify novel relationships, but also to investigate rare disease outcomes. However, the focus of the presented small case study is on the method and the type of analysis that can be enabled rather than on the actual medical associations. Therefore, the presented result should not be over-interpreted from a medical viewpoint.

Cardiovascular disease (CVD) is one of the major causes of death worldwide with an immense impact on public health. CVD is the general term for conditions affecting the heart and blood vessels including ischaemic heart disease and stroke. Understanding the risk factors driving the progression of CVD could help practitioners to intervene early or allow individuals to adjust lifestyle choices. To identify potential risk factors, we make use of the UK Biobank (UKB) \citep{sudlow_uk_2015}, a public large-scale biomedical database established in 2006 with a vast array of information on 502,628 participants recruited between 2006 and 2010. All participants were between 40-69 years of age at their recruitment date. 

In total, we have 393,464 participants with 1,594,586 interval observations and 6,880 MI events. Participants are left-truncated by the date of their entry and possibly right-censored, the latest at 2020-03-01. 
Additionally, we have 1,108 time-varying binary indicators for each ICD10 code representing if an individual ever had a corresponding disease at the corresponding point in time. Furthermore, we have 12 baseline risk factors (sex, alcohol, smoking, LDL, HDL, Triglyceride, BMI(3 indicators), Blood Pressure(3 indicators)) that are generally be considered to influence an individual's risk of CVD. 
Overall, there are 1,120 time-varying covariates and sex in our model with covariates values taken from either the most recent recruitment visit, follow up, or entry in the hospital admission records and are assumed to be constant in-between measurements.

We transform the timeline from real-time to age in days as age is a dominant and powerful predictor. In such a cases it can be advisable to use age as the preferable time axis. See \citet{andersen_analysis_2021} for general guidelines.

The outcome of the study is a myocardial infarction defined as an ICD10 code in I21-I24. Other CVD events like ischemic heart disease (I20-I25), cerebrovascular disease (I60-I69), cardiac arrest (I46), heart failure (I50), and transient cerebral ischaemic attack (G45) have been used additionally as either exclusion criterion before study entry or as a possible competing event, which we treat here as a censoring instance for simplicity. Of interest is the MI event one year ahead - the last year of observation prior to the event is removed - to avoid identifying factors that might be part of the diagnostic process itself. 

A more detailed description of the study can be found in Section 10 of the Supplementary materials.

For analysis, we use a similar setup as in the high-dimensional simulation in Section \ref{subsec:highdim}.
The prior distribution is $\theta_l \sim \mathrm{Student} {\shortminus} t(\nu=1, s=0.001)\,,$
where we use a low rank(50) Multivariate Normal as our approximation family for variational inference. 
The batch size for inference is 8,192 individuals.

\begin{figure}[h!]
    \centering
    \includegraphics[width=0.8\textwidth]{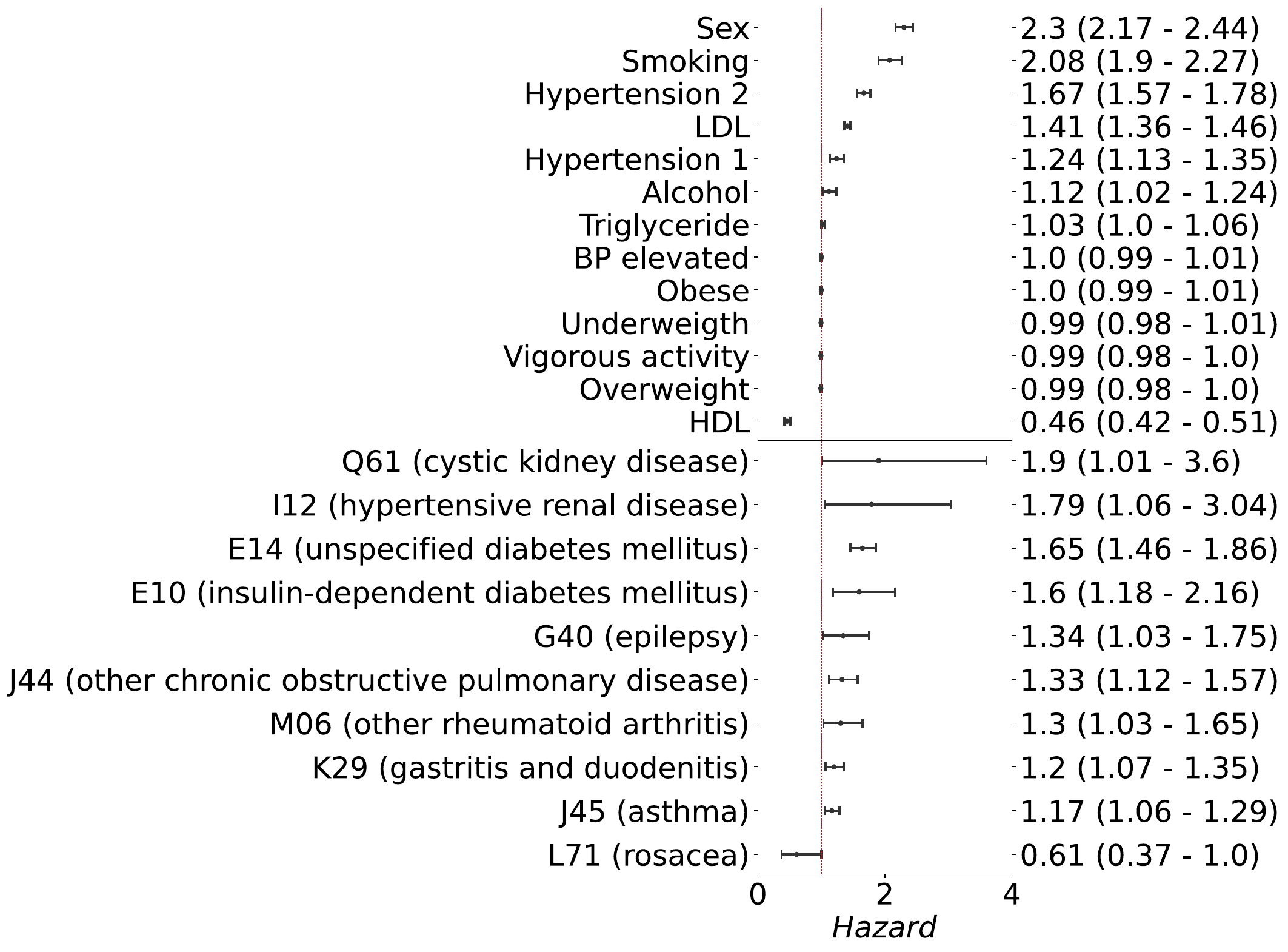}
    \caption{Forest plot for the hazard $\exp[\theta]$. The black dot represents the median parameter estimates with the corresponding $HPD_{95\%}$ around it. The upper panel shows the estimates for the baseline risk factors. We include all baseline risk estimates. The lower panel shows the parameter estimates for diseases with sufficient evidence of a non-zero effect based on the $HPD_{95\%}$ interval.}
    \label{fig:forest}
\end{figure}

Overall the results are in line with many of the reported associations in the literature.
The most relevant factors according to our estimation can be seen in Figure \ref{fig:forest}.

\citet{yusuf_modifiable_2020} report a hazard ratio of 1.74 (1.61 - 1.88) for diabetes without sub-categorization on CVD, which agrees with our estimates for unspecified diabetes 1.65 (1.46 - 1.86) and insulin-dependent diabetes 1.6 (1.18 - 2.16). \citet{millett_sex_2018} estimate sex-specific factors for smoking on MI as 3.46 (3.02 - 3.98) and 2.23 (2.03 - 2.44) for female and male, respectively, slightly higher then our estimate of 2.08 (1.9 - 2.27). \citet{mortensen_elevated_2020} estimate a hazard ratio of 1.34 (1.27 - 1.47) for LDL on MI, again close to our estimates of 1.41 (1.36 - 1.46). Chronic kidney disease has been identified by \citet{hippisley-cox_development_2017} as a risk factor for CVD with a hazard ratio of 2.09 (1.87 - 2.34) and 1.94 (1.72 - 2.19), for male and female, respectively. These estimates overlap with our cystic kidney disease and hypertensive renal disease estimates of 1.9 (1.01-3.6) and 1.79 (1.06 - 3.04), respectively. The same study also estimates a hazard ratio for rheumatoid arthritis of 1.24 (1.19 - 1.28), similar to our estimate of 1.30 (1.03 - 1.65). The association between rosacea and MI is less well established, however, \citet{egeberg_assessment_2016} estimates a relative risk ratio of 0.75\% (0.57 - 1.00) in line with our estimate of 0.61 (0.37 - 1.00).
The concordance index for the model on the train, valid, and test split are 0.72, 0.72, and 0.69, respectively.
This type of analysis could be further extended to include more of the available information in the UKB and refine associations, but more importantly, with the number of participants in these cohorts we can now reveal and study more nuanced outcomes and systematically evaluate multivariate effects. 

\section{Limitations and future directions}
\label{sec:lim}
The presented analysis mainly focused on the scalability of the Cox model in high-dimensional and large-scale datasets with time-varying covariates. Further research will be needed to extend the proposed method to some of the challenges that may arise when analyzing EHR. We provide a brief outline of some possible future directions of this research.

A limitation of the proposed analysis is the assumption of proportional hazards and the lack of means to assess. Developing efficient algorithms to scale residuals analysis to these vast amounts of data to evaluate model performance and identify possible deficiencies will be crucial. 

Further extending the flexibility of the method by non-linear and time-dependent covariate effects is another factor to consider. 
However, challenges may arise through the different dimensions on which the effects can act when considering time-varying covariates. The effect may vary by timepoint of measurement (early vs. late pregnancy) and by the progression in time (decaying or increasing effects with time).

While EHR provide a rich data source, the assumption that the measurements are constant in-between time points or are measured without error may introduce a bias. Modeling the longitudinal data jointly with event times would be an alternative approach, however, there are additional inferential and computational challenges.

Competing events naturally arise when analyzing EHR and pose challenges for the interpretation of results. In this study we focused on the cause-specific hazard, however, adopting different approaches like the Fine-Gray model, i.e. subdistribution hazard, may be preferable in some cases while challenging for internal time-varying covariates.

Using SVI rather than MCMC provides substantial compute time advantages at the cost of lacking guarantees for the approximation. 
The adoption of SVI for more complex models e.g. hierarchical models or frailty models imposes additional challenges as defining a good approximating family will require further research and evaluation.
The performance of our approach in the context of high-dimensional data is promising. Nevertheless, using the lower rank approximation to the Normal distribution as the variational family will require further theoretical justifications.
Extending SVI to very high-dimensional datasets e.g genetic association studies or using flexible variational families e.g. normalizing flows are promising research directions.

An open question, remains the optimal batch size to choose for a given application. Generally, we have shown that the method is reasonably robust for a variety of different batch sizes without incurring too much of a bias.

\section{Concluding remarks}
\label{sec:dis}

We have proposed a Bayesian version of Cox partial likelihood that can be used on high-dimensional and large-scale datasets with time-varying covariates. For the simulated datasets and optimization parameters, our method provides (almost) unbiased estimates with good uncertainty quantification while subsampling data. 
This approach enables researchers to extend the Cox model, one of the most widely applied methods in biomedical research, to new data sources like biobanks and EHR, the combination of which can provide new insights into our understanding of diseases.
Our implementation enables the joint analysis of tens of thousands of time-varying covariates for millions of individuals, providing a framework to utilize population-scale EHR.
Based on a counting process representation, our method can easily be extended to a wider variety of scenarios like recurrent events or more elaborate missing data patterns.
As shown in the case of myocardial infarction, we can replicate many of the results previously reported in the literature and characterize their joint effects. Extending the analysis by the additional data typically available in EHR or focusing on less well-studied disease could be an interesting avenue to follow.

%
%


\begin{funding}
AWJ and MG are supported by grant NNF17OC0027594 from the Novo Nordisk Foundation.
The data for the UK Biobank was accessed by application 45761.
No potential competing interest was reported by the authors.
\end{funding}

\begin{supplement}
\stitle{Supplemental Code and Data}
\sdescription{Python-package \textit{probcox}. Contains the code to fit the probabilistic cox regression method described in the article. Further contains all datasets and scripts (except the UKB data - additional fake data simulation + analysis are provided instead) used as examples in the article. Notebooks are provided to readily replicate all results, tables and figures from the paper \citep{jung2022}.
(zip file or \url{https://github.com/alexwjung/ProbCox})}
\end{supplement}

\begin{supplement}
\stitle{Supplemental Simulation and Application Results}
\sdescription{Contains additional tables and results described in the article \citep{jung2022}. 
(pdf file)}
\end{supplement}


\bibliographystyle{imsart-nameyear} 
\bibliography{bibliography}       

\begin{thebibliography}{51}

\bibitem[\protect\citeauthoryear{Alvares et~al.}{2021}]{alvares_bayesian_2021}
\begin{barticle}[author]
\bauthor{\bsnm{Alvares},~\bfnm{Danilo}\binits{D.}},
  \bauthor{\bsnm{Lázaro},~\bfnm{Elena}\binits{E.}},
  \bauthor{\bsnm{Gómez-Rubio},~\bfnm{Virgilio}\binits{V.}} \AND
  \bauthor{\bsnm{Armero},~\bfnm{Carmen}\binits{C.}}
(\byear{2021}).
\btitle{Bayesian survival analysis with {BUGS}}.
\bjournal{Statistics in Medicine}
\bvolume{40}
\bpages{2975--3020}.
\bnote{\_eprint: https://onlinelibrary.wiley.com/doi/pdf/10.1002/sim.8933}.
\bdoi{10.1002/sim.8933}
\end{barticle}
\endbibitem

\bibitem[\protect\citeauthoryear{Andersen and Gill}{1982}]{andersen_coxs_1982}
\begin{barticle}[author]
\bauthor{\bsnm{Andersen},~\bfnm{P.~K.}\binits{P.~K.}} \AND
  \bauthor{\bsnm{Gill},~\bfnm{R.~D.}\binits{R.~D.}}
(\byear{1982}).
\btitle{Cox's {Regression} {Model} for {Counting} {Processes}: {A} {Large}
  {Sample} {Study}}.
\bjournal{The Annals of Statistics}
\bvolume{10}
\bpages{1100--1120}.
\bdoi{10.1214/aos/1176345976}
\end{barticle}
\endbibitem

\bibitem[\protect\citeauthoryear{Andersen
  et~al.}{1993}]{andersen_statistical_1993}
\begin{bbook}[author]
\bauthor{\bsnm{Andersen},~\bfnm{Per~Kragh}\binits{P.~K.}},
  \bauthor{\bsnm{Borgan},~\bfnm{Ornulf}\binits{O.}},
  \bauthor{\bsnm{Gill},~\bfnm{Richard~D.}\binits{R.~D.}} \AND
  \bauthor{\bsnm{Keiding},~\bfnm{Niels}\binits{N.}}
(\byear{1993}).
\btitle{Statistical {Models} {Based} on {Counting} {Processes}}.
\bseries{Springer {Series} in {Statistics}}.
\bpublisher{Springer-Verlag}, \baddress{New York}.
\bdoi{10.1007/978-1-4612-4348-9}
\end{bbook}
\endbibitem

\bibitem[\protect\citeauthoryear{Andersen
  et~al.}{2021}]{andersen_analysis_2021}
\begin{barticle}[author]
\bauthor{\bsnm{Andersen},~\bfnm{Per~Kragh}\binits{P.~K.}},
  \bauthor{\bsnm{Perme},~\bfnm{Maja~Pohar}\binits{M.~P.}},
  \bauthor{\bsnm{Houwelingen},~\bfnm{Hans C.~van}\binits{H.~C.~v.}},
  \bauthor{\bsnm{Cook},~\bfnm{Richard~J.}\binits{R.~J.}},
  \bauthor{\bsnm{Joly},~\bfnm{Pierre}\binits{P.}},
  \bauthor{\bsnm{Martinussen},~\bfnm{Torben}\binits{T.}},
  \bauthor{\bsnm{Taylor},~\bfnm{Jeremy M.~G.}\binits{J.~M.~G.}},
  \bauthor{\bsnm{Abrahamowicz},~\bfnm{Michal}\binits{M.}} \AND
  \bauthor{\bsnm{Therneau},~\bfnm{Terry~M.}\binits{T.~M.}}
(\byear{2021}).
\btitle{Analysis of time-to-event for observational studies: {Guidance} to the
  use of intensity models}.
\bjournal{Statistics in Medicine}
\bvolume{40}
\bpages{185--211}.
\bnote{\_eprint: https://onlinelibrary.wiley.com/doi/pdf/10.1002/sim.8757}.
\bdoi{https://doi.org/10.1002/sim.8757}
\end{barticle}
\endbibitem

\bibitem[\protect\citeauthoryear{Breheny and
  Huang}{2011}]{breheny_coordinate_2011}
\begin{barticle}[author]
\bauthor{\bsnm{Breheny},~\bfnm{Patrick}\binits{P.}} \AND
  \bauthor{\bsnm{Huang},~\bfnm{Jian}\binits{J.}}
(\byear{2011}).
\btitle{Coordinate descent algorithms for nonconvex penalized regression, with
  applications to biological feature selection}.
\bjournal{The Annals of Applied Statistics}
\bvolume{5}
\bpages{232--253}.
\bnote{Publisher: Institute of Mathematical Statistics}.
\bdoi{10.1214/10-AOAS388}
\end{barticle}
\endbibitem

\bibitem[\protect\citeauthoryear{Clift et~al.}{2020}]{clift_living_2020}
\begin{barticle}[author]
\bauthor{\bsnm{Clift},~\bfnm{Ash~K.}\binits{A.~K.}},
  \bauthor{\bsnm{Coupland},~\bfnm{Carol A.~C.}\binits{C.~A.~C.}},
  \bauthor{\bsnm{Keogh},~\bfnm{Ruth~H.}\binits{R.~H.}},
  \bauthor{\bsnm{Diaz-Ordaz},~\bfnm{Karla}\binits{K.}},
  \bauthor{\bsnm{Williamson},~\bfnm{Elizabeth}\binits{E.}},
  \bauthor{\bsnm{Harrison},~\bfnm{Ewen~M.}\binits{E.~M.}},
  \bauthor{\bsnm{Hayward},~\bfnm{Andrew}\binits{A.}},
  \bauthor{\bsnm{Hemingway},~\bfnm{Harry}\binits{H.}},
  \bauthor{\bsnm{Horby},~\bfnm{Peter}\binits{P.}},
  \bauthor{\bsnm{Mehta},~\bfnm{Nisha}\binits{N.}},
  \bauthor{\bsnm{Benger},~\bfnm{Jonathan}\binits{J.}},
  \bauthor{\bsnm{Khunti},~\bfnm{Kamlesh}\binits{K.}},
  \bauthor{\bsnm{Spiegelhalter},~\bfnm{David}\binits{D.}},
  \bauthor{\bsnm{Sheikh},~\bfnm{Aziz}\binits{A.}},
  \bauthor{\bsnm{Valabhji},~\bfnm{Jonathan}\binits{J.}},
  \bauthor{\bsnm{Lyons},~\bfnm{Ronan~A.}\binits{R.~A.}},
  \bauthor{\bsnm{Robson},~\bfnm{John}\binits{J.}},
  \bauthor{\bsnm{Semple},~\bfnm{Malcolm~G.}\binits{M.~G.}},
  \bauthor{\bsnm{Kee},~\bfnm{Frank}\binits{F.}},
  \bauthor{\bsnm{Johnson},~\bfnm{Peter}\binits{P.}},
  \bauthor{\bsnm{Jebb},~\bfnm{Susan}\binits{S.}},
  \bauthor{\bsnm{Williams},~\bfnm{Tony}\binits{T.}} \AND
  \bauthor{\bsnm{Hippisley-Cox},~\bfnm{Julia}\binits{J.}}
(\byear{2020}).
\btitle{Living risk prediction algorithm ({QCOVID}) for risk of hospital
  admission and mortality from coronavirus 19 in adults: national derivation
  and validation cohort study}.
\bjournal{BMJ (Clinical research ed.)}
\bvolume{371}
\bpages{m3731}.
\bdoi{10.1136/bmj.m3731}
\end{barticle}
\endbibitem

\bibitem[\protect\citeauthoryear{Cox}{1972}]{cox_regression_1972}
\begin{barticle}[author]
\bauthor{\bsnm{Cox},~\bfnm{D.~R.}\binits{D.~R.}}
(\byear{1972}).
\btitle{Regression {Models} and {Life}-{Tables}}.
\bjournal{Journal of the Royal Statistical Society. Series B (Methodological)}
\bvolume{34}
\bpages{187--220}.
\bnote{Publisher: [Royal Statistical Society, Wiley]}.
\end{barticle}
\endbibitem

\bibitem[\protect\citeauthoryear{Cox}{1975}]{cox_partial_1975}
\begin{barticle}[author]
\bauthor{\bsnm{Cox},~\bfnm{D.~R.}\binits{D.~R.}}
(\byear{1975}).
\btitle{Partial {Likelihood}}.
\bjournal{Biometrika}
\bvolume{62}
\bpages{269--276}.
\bnote{Publisher: [Oxford University Press, Biometrika Trust]}.
\bdoi{10.2307/2335362}
\end{barticle}
\endbibitem

\bibitem[\protect\citeauthoryear{Dawber, Meadors and
  Moore}{1951}]{dawber_epidemiological_1951}
\begin{barticle}[author]
\bauthor{\bsnm{Dawber},~\bfnm{T.~R.}\binits{T.~R.}},
  \bauthor{\bsnm{Meadors},~\bfnm{G.~F.}\binits{G.~F.}} \AND
  \bauthor{\bsnm{Moore},~\bfnm{F.~E.}\binits{F.~E.}}
(\byear{1951}).
\btitle{Epidemiological approaches to heart disease: the {Framingham} {Study}}.
\bjournal{American Journal of Public Health and the Nation's Health}
\bvolume{41}
\bpages{279--281}.
\bdoi{10.2105/ajph.41.3.279}
\end{barticle}
\endbibitem

\bibitem[\protect\citeauthoryear{Egeberg
  et~al.}{2016}]{egeberg_assessment_2016}
\begin{barticle}[author]
\bauthor{\bsnm{Egeberg},~\bfnm{Alexander}\binits{A.}},
  \bauthor{\bsnm{Hansen},~\bfnm{Peter~R.}\binits{P.~R.}},
  \bauthor{\bsnm{Gislason},~\bfnm{Gunnar~H.}\binits{G.~H.}} \AND
  \bauthor{\bsnm{Thyssen},~\bfnm{Jacob~P.}\binits{J.~P.}}
(\byear{2016}).
\btitle{Assessment of the risk of cardiovascular disease in patients with
  rosacea}.
\bjournal{Journal of the American Academy of Dermatology}
\bvolume{75}
\bpages{336--339}.
\bdoi{10.1016/j.jaad.2016.02.1158}
\end{barticle}
\endbibitem

\bibitem[\protect\citeauthoryear{Friedman, Hastie and
  Tibshirani}{2010}]{friedman_regularization_2010}
\begin{barticle}[author]
\bauthor{\bsnm{Friedman},~\bfnm{Jerome}\binits{J.}},
  \bauthor{\bsnm{Hastie},~\bfnm{Trevor}\binits{T.}} \AND
  \bauthor{\bsnm{Tibshirani},~\bfnm{Rob}\binits{R.}}
(\byear{2010}).
\btitle{Regularization {Paths} for {Generalized} {Linear} {Models} via
  {Coordinate} {Descent}}.
\bjournal{Journal of Statistical Software}
\bvolume{33}
\bpages{1--22}.
\end{barticle}
\endbibitem

\bibitem[\protect\citeauthoryear{Hans}{2009}]{hans2009bayesian}
\begin{barticle}[author]
\bauthor{\bsnm{Hans},~\bfnm{Chris}\binits{C.}}
(\byear{2009}).
\btitle{Bayesian lasso regression}.
\bjournal{Biometrika}
\bvolume{96}
\bpages{835--845}.
\end{barticle}
\endbibitem

\bibitem[\protect\citeauthoryear{Hippisley-Cox, Coupland and
  Brindle}{2017}]{hippisley-cox_development_2017}
\begin{barticle}[author]
\bauthor{\bsnm{Hippisley-Cox},~\bfnm{Julia}\binits{J.}},
  \bauthor{\bsnm{Coupland},~\bfnm{Carol}\binits{C.}} \AND
  \bauthor{\bsnm{Brindle},~\bfnm{Peter}\binits{P.}}
(\byear{2017}).
\btitle{Development and validation of {QRISK3} risk prediction algorithms to
  estimate future risk of cardiovascular disease: prospective cohort study}.
\bjournal{BMJ}
\bvolume{357}
\bpages{j2099}.
\bnote{Publisher: British Medical Journal Publishing Group Section: Research}.
\bdoi{10.1136/bmj.j2099}
\end{barticle}
\endbibitem

\bibitem[\protect\citeauthoryear{Hippisley-Cox and
  Coupland}{2021}]{hippisley-cox_predicting_2021}
\begin{barticle}[author]
\bauthor{\bsnm{Hippisley-Cox},~\bfnm{Julia}\binits{J.}} \AND
  \bauthor{\bsnm{Coupland},~\bfnm{Carol}\binits{C.}}
(\byear{2021}).
\btitle{Predicting the risk of prostate cancer in asymptomatic men: a cohort
  study to develop and validate a novel algorithm}.
\bjournal{The British Journal of General Practice: The Journal of the Royal
  College of General Practitioners}
\bvolume{71}
\bpages{e364--e371}.
\bdoi{10.3399/bjgp20X714137}
\end{barticle}
\endbibitem

\bibitem[\protect\citeauthoryear{Hjort}{1990}]{hjort_nonparametric_1990}
\begin{barticle}[author]
\bauthor{\bsnm{Hjort},~\bfnm{Nils~Lid}\binits{N.~L.}}
(\byear{1990}).
\btitle{Nonparametric {Bayes} {Estimators} {Based} on {Beta} {Processes} in
  {Models} for {Life} {History} {Data}}.
\bjournal{The Annals of Statistics}
\bvolume{18}
\bpages{1259--1294}.
\bnote{Publisher: Institute of Mathematical Statistics}.
\end{barticle}
\endbibitem

\bibitem[\protect\citeauthoryear{Ibrahim, Chen and
  Sinha}{2001}]{ibrahim_bayesian_2001}
\begin{bbook}[author]
\bauthor{\bsnm{Ibrahim},~\bfnm{Joseph~G.}\binits{J.~G.}},
  \bauthor{\bsnm{Chen},~\bfnm{Ming-Hui}\binits{M.-H.}} \AND
  \bauthor{\bsnm{Sinha},~\bfnm{Debajyoti}\binits{D.}}
(\byear{2001}).
\btitle{Bayesian {Survival} {Analysis}}.
\bseries{Springer {Series} in {Statistics}}.
\bpublisher{Springer-Verlag}, \baddress{New York}.
\bdoi{10.1007/978-1-4757-3447-8}
\end{bbook}
\endbibitem

\bibitem[\protect\citeauthoryear{Jung and Gerstung}{2022}]{jung2022}
\begin{barticle}[author]
\bauthor{\bsnm{Jung},~\bfnm{Alexander~W.}\binits{A.~W.}} \AND
  \bauthor{\bsnm{Gerstung},~\bfnm{Moritz}\binits{M.}}
(\byear{2022}).
\btitle{Supplement to "Bayesian Cox Regression for Large-scale Inference with
  Applications to Electronic Health Records"}.
\bjournal{Annals of Applied Statistics}.
\end{barticle}
\endbibitem

\bibitem[\protect\citeauthoryear{Kalbfleisch}{1978}]{kalbfleisch_non-parametric_1978}
\begin{barticle}[author]
\bauthor{\bsnm{Kalbfleisch},~\bfnm{John~D.}\binits{J.~D.}}
(\byear{1978}).
\btitle{Non-{Parametric} {Bayesian} {Analysis} of {Survival} {Time} {Data}}.
\bjournal{Journal of the Royal Statistical Society. Series B (Methodological)}
\bvolume{40}
\bpages{214--221}.
\bnote{Publisher: [Royal Statistical Society, Wiley]}.
\end{barticle}
\endbibitem

\bibitem[\protect\citeauthoryear{Kalbfleisch and
  Prentice}{1973}]{kalbfleisch_marginal_1973}
\begin{barticle}[author]
\bauthor{\bsnm{Kalbfleisch},~\bfnm{J.~D.}\binits{J.~D.}} \AND
  \bauthor{\bsnm{Prentice},~\bfnm{R.~L.}\binits{R.~L.}}
(\byear{1973}).
\btitle{Marginal {Likelihoods} {Based} on {Cox}'s {Regression} and {Life}
  {Model}}.
\bjournal{Biometrika}
\bvolume{60}
\bpages{267--278}.
\bnote{Publisher: [Oxford University Press, Biometrika Trust]}.
\bdoi{10.2307/2334538}
\end{barticle}
\endbibitem

\bibitem[\protect\citeauthoryear{Kucukelbir
  et~al.}{2017}]{kucukelbir_automatic_2017}
\begin{barticle}[author]
\bauthor{\bsnm{Kucukelbir},~\bfnm{Alp}\binits{A.}},
  \bauthor{\bsnm{Tran},~\bfnm{Dustin}\binits{D.}},
  \bauthor{\bsnm{Ranganath},~\bfnm{Rajesh}\binits{R.}},
  \bauthor{\bsnm{Gelman},~\bfnm{Andrew}\binits{A.}} \AND
  \bauthor{\bsnm{Blei},~\bfnm{David~M.}\binits{D.~M.}}
(\byear{2017}).
\btitle{Automatic {Differentiation} {Variational} {Inference}}.
\bjournal{Journal of Machine Learning Research}
\bvolume{18}
\bpages{1--45}.
\end{barticle}
\endbibitem

\bibitem[\protect\citeauthoryear{Kvamme, Borgan and
  Scheel}{2019}]{kvamme_time--event_2019}
\begin{barticle}[author]
\bauthor{\bsnm{Kvamme},~\bfnm{Havard}\binits{H.}},
  \bauthor{\bsnm{Borgan},~\bfnm{Ornulf}\binits{O.}} \AND
  \bauthor{\bsnm{Scheel},~\bfnm{Ida}\binits{I.}}
(\byear{2019}).
\btitle{Time-to-{Event} {Prediction} with {Neural} {Networks} and {Cox}
  {Regression}}.
\bjournal{Journal of Machine Learning Research}
\bvolume{20}
\bpages{1--30}.
\end{barticle}
\endbibitem

\bibitem[\protect\citeauthoryear{Laud, Damien and
  Smith}{1998}]{laud_bayesian_1998}
\begin{bincollection}[author]
\bauthor{\bsnm{Laud},~\bfnm{Purushottam~W.}\binits{P.~W.}},
  \bauthor{\bsnm{Damien},~\bfnm{Paul}\binits{P.}} \AND
  \bauthor{\bsnm{Smith},~\bfnm{Adrian F.~M.}\binits{A.~F.~M.}}
(\byear{1998}).
\btitle{Bayesian {Nonparametric} and {Covariate} {Analysis} of {Failure} {Time}
  {Data}}.
In \bbooktitle{Practical {Nonparametric} and {Semiparametric} {Bayesian}
  {Statistics}},
(\beditor{\bfnm{Dipak}\binits{D.}~\bsnm{Dey}},
  \beditor{\bfnm{Peter}\binits{P.}~\bsnm{Müller}} \AND
  \beditor{\bfnm{Debajyoti}\binits{D.}~\bsnm{Sinha}}, eds.).
\bseries{Lecture {Notes} in {Statistics}}
\bpages{213--225}.
\bpublisher{Springer}, \baddress{New York, NY}.
\end{bincollection}
\endbibitem

\bibitem[\protect\citeauthoryear{Lewandowski, Kurowicka and
  Joe}{2009}]{lewandowski2009generating}
\begin{barticle}[author]
\bauthor{\bsnm{Lewandowski},~\bfnm{Daniel}\binits{D.}},
  \bauthor{\bsnm{Kurowicka},~\bfnm{Dorota}\binits{D.}} \AND
  \bauthor{\bsnm{Joe},~\bfnm{Harry}\binits{H.}}
(\byear{2009}).
\btitle{Generating random correlation matrices based on vines and extended
  onion method}.
\bjournal{Journal of multivariate analysis}
\bvolume{100}
\bpages{1989--2001}.
\end{barticle}
\endbibitem

\bibitem[\protect\citeauthoryear{Li et~al.}{2020}]{li_fast_2020}
\begin{barticle}[author]
\bauthor{\bsnm{Li},~\bfnm{Ruilin}\binits{R.}},
  \bauthor{\bsnm{Chang},~\bfnm{Christopher}\binits{C.}},
  \bauthor{\bsnm{Justesen},~\bfnm{Johanne~M}\binits{J.~M.}},
  \bauthor{\bsnm{Tanigawa},~\bfnm{Yosuke}\binits{Y.}},
  \bauthor{\bsnm{Qiang},~\bfnm{Junyang}\binits{J.}},
  \bauthor{\bsnm{Hastie},~\bfnm{Trevor}\binits{T.}},
  \bauthor{\bsnm{Rivas},~\bfnm{Manuel~A}\binits{M.~A.}} \AND
  \bauthor{\bsnm{Tibshirani},~\bfnm{Robert}\binits{R.}}
(\byear{2020}).
\btitle{Fast {Lasso} method for large-scale and ultrahigh-dimensional {Cox}
  model with applications to {UK} {Biobank}}.
\bjournal{Biostatistics}
\bvolume{kxaa038}.
\bdoi{10.1093/biostatistics/kxaa038}
\end{barticle}
\endbibitem

\bibitem[\protect\citeauthoryear{Millett, Peters and
  Woodward}{2018}]{millett_sex_2018}
\begin{barticle}[author]
\bauthor{\bsnm{Millett},~\bfnm{Elizabeth R~C}\binits{E.~R.~C.}},
  \bauthor{\bsnm{Peters},~\bfnm{Sanne A~E}\binits{S.~A.~E.}} \AND
  \bauthor{\bsnm{Woodward},~\bfnm{Mark}\binits{M.}}
(\byear{2018}).
\btitle{Sex differences in risk factors for myocardial infarction: cohort study
  of {UK} {Biobank} participants}.
\bjournal{BMJ}
\bpages{k4247}.
\bdoi{10.1136/bmj.k4247}
\end{barticle}
\endbibitem

\bibitem[\protect\citeauthoryear{Mittal
  et~al.}{2014}]{mittal_high-dimensional_2014}
\begin{barticle}[author]
\bauthor{\bsnm{Mittal},~\bfnm{S.}\binits{S.}},
  \bauthor{\bsnm{Madigan},~\bfnm{D.}\binits{D.}},
  \bauthor{\bsnm{Burd},~\bfnm{R.~S.}\binits{R.~S.}} \AND
  \bauthor{\bsnm{Suchard},~\bfnm{M.~A.}\binits{M.~A.}}
(\byear{2014}).
\btitle{High-dimensional, massive sample-size {Cox} proportional hazards
  regression for survival analysis}.
\bjournal{Biostatistics}
\bvolume{15}
\bpages{207--221}.
\bdoi{10.1093/biostatistics/kxt043}
\end{barticle}
\endbibitem

\bibitem[\protect\citeauthoryear{Mohamed et~al.}{2020}]{mohamed_monte_2019}
\begin{barticle}[author]
\bauthor{\bsnm{Mohamed},~\bfnm{Shakir}\binits{S.}},
  \bauthor{\bsnm{Rosca},~\bfnm{Mihaela}\binits{M.}},
  \bauthor{\bsnm{Figurnov},~\bfnm{Michael}\binits{M.}} \AND
  \bauthor{\bsnm{Mnih},~\bfnm{Andriy}\binits{A.}}
(\byear{2020}).
\btitle{Monte Carlo Gradient Estimation in Machine Learning.}
\bjournal{J. Mach. Learn. Res.}
\bvolume{21}
\bpages{1--62}.
\end{barticle}
\endbibitem

\bibitem[\protect\citeauthoryear{Mortensen and
  Nordestgaard}{2020}]{mortensen_elevated_2020}
\begin{barticle}[author]
\bauthor{\bsnm{Mortensen},~\bfnm{Martin~Bødtker}\binits{M.~B.}} \AND
  \bauthor{\bsnm{Nordestgaard},~\bfnm{Børge~Grønne}\binits{B.~G.}}
(\byear{2020}).
\btitle{Elevated {LDL} cholesterol and increased risk of myocardial infarction
  and atherosclerotic cardiovascular disease in individuals aged 70–100
  years: a contemporary primary prevention cohort}.
\bjournal{The Lancet}
\bvolume{396}
\bpages{1644--1652}.
\bdoi{10.1016/S0140-6736(20)32233-9}
\end{barticle}
\endbibitem

\bibitem[\protect\citeauthoryear{Nikooienejad, Wang and
  Johnson}{2020}]{nikooienejad_bayesian_2020}
\begin{barticle}[author]
\bauthor{\bsnm{Nikooienejad},~\bfnm{Amir}\binits{A.}},
  \bauthor{\bsnm{Wang},~\bfnm{Wenyi}\binits{W.}} \AND
  \bauthor{\bsnm{Johnson},~\bfnm{Valen~E.}\binits{V.~E.}}
(\byear{2020}).
\btitle{Bayesian variable selection for survival data using inverse moment
  priors}.
\bjournal{The Annals of Applied Statistics}
\bvolume{14}
\bpages{809--828}.
\bnote{Publisher: Institute of Mathematical Statistics}.
\bdoi{10.1214/20-AOAS1325}
\end{barticle}
\endbibitem

\bibitem[\protect\citeauthoryear{Park and Casella}{2008}]{park_bayesian_2008}
\begin{barticle}[author]
\bauthor{\bsnm{Park},~\bfnm{Trevor}\binits{T.}} \AND
  \bauthor{\bsnm{Casella},~\bfnm{George}\binits{G.}}
(\byear{2008}).
\btitle{The {Bayesian} {Lasso}}.
\bjournal{Journal of the American Statistical Association}
\bvolume{103}
\bpages{681--686}.
\bnote{Publisher: Taylor \& Francis \_eprint:
  https://doi.org/10.1198/016214508000000337}.
\bdoi{10.1198/016214508000000337}
\end{barticle}
\endbibitem

\bibitem[\protect\citeauthoryear{Qiou, Ravishanker and
  Dey}{1999}]{qiou_multivariate_1999}
\begin{barticle}[author]
\bauthor{\bsnm{Qiou},~\bfnm{Z.}\binits{Z.}},
  \bauthor{\bsnm{Ravishanker},~\bfnm{N.}\binits{N.}} \AND
  \bauthor{\bsnm{Dey},~\bfnm{D.~K.}\binits{D.~K.}}
(\byear{1999}).
\btitle{Multivariate survival analysis with positive stable frailties}.
\bjournal{Biometrics}
\bvolume{55}
\bpages{637--644}.
\bdoi{10.1111/j.0006-341x.1999.00637.x}
\end{barticle}
\endbibitem

\bibitem[\protect\citeauthoryear{Ranganath, Gerrish and
  Blei}{2014}]{ranganath_black_2014}
\begin{binproceedings}[author]
\bauthor{\bsnm{Ranganath},~\bfnm{Rajesh}\binits{R.}},
  \bauthor{\bsnm{Gerrish},~\bfnm{Sean}\binits{S.}} \AND
  \bauthor{\bsnm{Blei},~\bfnm{David}\binits{D.}}
(\byear{2014}).
\btitle{Black box variational inference}.
In \bbooktitle{Artificial intelligence and statistics}
\bpages{814--822}.
\bpublisher{PMLR}.
\end{binproceedings}
\endbibitem

\bibitem[\protect\citeauthoryear{Sharef et~al.}{2010}]{sharef_bayesian_2010}
\begin{barticle}[author]
\bauthor{\bsnm{Sharef},~\bfnm{Emmanuel}\binits{E.}},
  \bauthor{\bsnm{Strawderman},~\bfnm{Robert~L.}\binits{R.~L.}},
  \bauthor{\bsnm{Ruppert},~\bfnm{David}\binits{D.}},
  \bauthor{\bsnm{Cowen},~\bfnm{Mark}\binits{M.}} \AND
  \bauthor{\bsnm{Halasyamani},~\bfnm{Lakshmi}\binits{L.}}
(\byear{2010}).
\btitle{Bayesian adaptive {B}-spline estimation in proportional hazards frailty
  models}.
\bjournal{Electronic Journal of Statistics}
\bvolume{4}
\bpages{606--642}.
\bnote{Publisher: Institute of Mathematical Statistics and Bernoulli Society}.
\bdoi{10.1214/10-EJS566}
\end{barticle}
\endbibitem

\bibitem[\protect\citeauthoryear{Shin, Bhattacharya and
  Johnson}{2018}]{shin_scalable_2018}
\begin{barticle}[author]
\bauthor{\bsnm{Shin},~\bfnm{Minsuk}\binits{M.}},
  \bauthor{\bsnm{Bhattacharya},~\bfnm{Anirban}\binits{A.}} \AND
  \bauthor{\bsnm{Johnson},~\bfnm{Valen~E}\binits{V.~E.}}
(\byear{2018}).
\btitle{Scalable {Bayesian} variable selection using nonlocal prior densities
  in ultrahigh-dimensional settings}.
\bjournal{Statistica Sinica}
\bvolume{28}
\bpages{1053}.
\bnote{Publisher: NIH Public Access}.
\end{barticle}
\endbibitem

\bibitem[\protect\citeauthoryear{Simon
  et~al.}{2011}]{simon_regularization_2011}
\begin{barticle}[author]
\bauthor{\bsnm{Simon},~\bfnm{Noah}\binits{N.}},
  \bauthor{\bsnm{Friedman},~\bfnm{Jerome}\binits{J.}},
  \bauthor{\bsnm{Hastie},~\bfnm{Trevor}\binits{T.}} \AND
  \bauthor{\bsnm{Tibshirani},~\bfnm{Robert}\binits{R.}}
(\byear{2011}).
\btitle{Regularization {Paths} for {Cox}'s {Proportional} {Hazards} {Model} via
  {Coordinate} {Descent}}.
\bjournal{Journal of Statistical Software}
\bvolume{39}
\bpages{1--13}.
\bdoi{10.18637/jss.v039.i05}
\end{barticle}
\endbibitem

\bibitem[\protect\citeauthoryear{Sinha}{1993}]{sinha_semiparametric_1993}
\begin{barticle}[author]
\bauthor{\bsnm{Sinha},~\bfnm{Debajyoti}\binits{D.}}
(\byear{1993}).
\btitle{Semiparametric {Bayesian} {Analysis} of {Multiple} {Event} {Time}
  {Data}}.
\bjournal{Journal of the American Statistical Association}
\bvolume{88}
\bpages{979--983}.
\bnote{Publisher: [American Statistical Association, Taylor \& Francis, Ltd.]}.
\bdoi{10.2307/2290789}
\end{barticle}
\endbibitem

\bibitem[\protect\citeauthoryear{Sinha, Ibrahim and
  Chen}{2003}]{sinha_bayesian_2003}
\begin{barticle}[author]
\bauthor{\bsnm{Sinha},~\bfnm{Debajyoti}\binits{D.}},
  \bauthor{\bsnm{Ibrahim},~\bfnm{Joseph~G.}\binits{J.~G.}} \AND
  \bauthor{\bsnm{Chen},~\bfnm{Ming-Hui}\binits{M.-H.}}
(\byear{2003}).
\btitle{A {Bayesian} {Justification} of {Cox}'s {Partial} {Likelihood}}.
\bjournal{Biometrika}
\bvolume{90}
\bpages{629--641}.
\bnote{Publisher: [Oxford University Press, Biometrika Trust]}.
\end{barticle}
\endbibitem

\bibitem[\protect\citeauthoryear{Sudlow et~al.}{2015}]{sudlow_uk_2015}
\begin{barticle}[author]
\bauthor{\bsnm{Sudlow},~\bfnm{Cathie}\binits{C.}},
  \bauthor{\bsnm{Gallacher},~\bfnm{John}\binits{J.}},
  \bauthor{\bsnm{Allen},~\bfnm{Naomi}\binits{N.}},
  \bauthor{\bsnm{Beral},~\bfnm{Valerie}\binits{V.}},
  \bauthor{\bsnm{Burton},~\bfnm{Paul}\binits{P.}},
  \bauthor{\bsnm{Danesh},~\bfnm{John}\binits{J.}},
  \bauthor{\bsnm{Downey},~\bfnm{Paul}\binits{P.}},
  \bauthor{\bsnm{Elliott},~\bfnm{Paul}\binits{P.}},
  \bauthor{\bsnm{Green},~\bfnm{Jane}\binits{J.}},
  \bauthor{\bsnm{Landray},~\bfnm{Martin}\binits{M.}},
  \bauthor{\bsnm{Liu},~\bfnm{Bette}\binits{B.}},
  \bauthor{\bsnm{Matthews},~\bfnm{Paul}\binits{P.}},
  \bauthor{\bsnm{Ong},~\bfnm{Giok}\binits{G.}},
  \bauthor{\bsnm{Pell},~\bfnm{Jill}\binits{J.}},
  \bauthor{\bsnm{Silman},~\bfnm{Alan}\binits{A.}},
  \bauthor{\bsnm{Young},~\bfnm{Alan}\binits{A.}},
  \bauthor{\bsnm{Sprosen},~\bfnm{Tim}\binits{T.}},
  \bauthor{\bsnm{Peakman},~\bfnm{Tim}\binits{T.}} \AND
  \bauthor{\bsnm{Collins},~\bfnm{Rory}\binits{R.}}
(\byear{2015}).
\btitle{{UK} biobank: an open access resource for identifying the causes of a
  wide range of complex diseases of middle and old age}.
\bjournal{PLoS medicine}
\bvolume{12}
\bpages{e1001779}.
\bdoi{10.1371/journal.pmed.1001779}
\end{barticle}
\endbibitem

\bibitem[\protect\citeauthoryear{Sylvestre and
  Abrahamowicz}{2008}]{sylvestre_comparison_2008}
\begin{barticle}[author]
\bauthor{\bsnm{Sylvestre},~\bfnm{Marie-Pierre}\binits{M.-P.}} \AND
  \bauthor{\bsnm{Abrahamowicz},~\bfnm{Michal}\binits{M.}}
(\byear{2008}).
\btitle{Comparison of algorithms to generate event times conditional on
  time-dependent covariates}.
\bjournal{Statistics in Medicine}
\bvolume{27}
\bpages{2618--2634}.
\bnote{\_eprint: https://onlinelibrary.wiley.com/doi/pdf/10.1002/sim.3092}.
\bdoi{https://doi.org/10.1002/sim.3092}
\end{barticle}
\endbibitem

\bibitem[\protect\citeauthoryear{Tarkhan and
  Simon}{2020}]{tarkhan_bigsurvsgd_2020}
\begin{barticle}[author]
\bauthor{\bsnm{Tarkhan},~\bfnm{Aliasghar}\binits{A.}} \AND
  \bauthor{\bsnm{Simon},~\bfnm{Noah}\binits{N.}}
(\byear{2020}).
\btitle{{BigSurvSGD}: {Big} {Survival} {Data} {Analysis} via {Stochastic}
  {Gradient} {Descent}}.
\bjournal{arXiv:2003.00116 [math, stat]}.
\bnote{arXiv: 2003.00116}.
\end{barticle}
\endbibitem

\bibitem[\protect\citeauthoryear{Therneau}{2021}]{therneau_package_2021}
\begin{bbook}[author]
\bauthor{\bsnm{Therneau},~\bfnm{Terry~M.}\binits{T.~M.}}
(\byear{2021}).
\btitle{A {Package} for {Survival} {Analysis} in {R}}.
\end{bbook}
\endbibitem

\bibitem[\protect\citeauthoryear{Therneau and
  Grambsch}{2000}]{therneau_cox_2000}
\begin{bincollection}[author]
\bauthor{\bsnm{Therneau},~\bfnm{Terry~M}\binits{T.~M.}} \AND
  \bauthor{\bsnm{Grambsch},~\bfnm{Patricia~M}\binits{P.~M.}}
(\byear{2000}).
\btitle{The cox model}.
In \bbooktitle{Modeling survival data: extending the {Cox} model}
\bpages{39--77}.
\bpublisher{Springer}.
\end{bincollection}
\endbibitem

\bibitem[\protect\citeauthoryear{Tibshirani}{1996}]{tibshirani_regression_1996}
\begin{barticle}[author]
\bauthor{\bsnm{Tibshirani},~\bfnm{Robert}\binits{R.}}
(\byear{1996}).
\btitle{Regression {Shrinkage} and {Selection} via the {Lasso}}.
\bjournal{Journal of the Royal Statistical Society. Series B (Methodological)}
\bvolume{58}
\bpages{267--288}.
\bnote{Publisher: [Royal Statistical Society, Wiley]}.
\end{barticle}
\endbibitem

\bibitem[\protect\citeauthoryear{Tibshirani}{1997}]{tibshirani_lasso_1997}
\begin{barticle}[author]
\bauthor{\bsnm{Tibshirani},~\bfnm{R.}\binits{R.}}
(\byear{1997}).
\btitle{The lasso method for variable selection in the {Cox} model}.
\bjournal{Statistics in Medicine}
\bvolume{16}
\bpages{385--395}.
\bdoi{10.1002/(sici)1097-0258(19970228)16:4<385::aid-sim380>3.0.co;2-3}
\end{barticle}
\endbibitem

\bibitem[\protect\citeauthoryear{Wang et~al.}{2021}]{wang2021fast}
\begin{barticle}[author]
\bauthor{\bsnm{Wang},~\bfnm{Yan}\binits{Y.}},
  \bauthor{\bsnm{Hong},~\bfnm{Chuan}\binits{C.}},
  \bauthor{\bsnm{Palmer},~\bfnm{Nathan}\binits{N.}},
  \bauthor{\bsnm{Di},~\bfnm{Qian}\binits{Q.}},
  \bauthor{\bsnm{Schwartz},~\bfnm{Joel}\binits{J.}},
  \bauthor{\bsnm{Kohane},~\bfnm{Isaac}\binits{I.}} \AND
  \bauthor{\bsnm{Cai},~\bfnm{Tianxi}\binits{T.}}
(\byear{2021}).
\btitle{A fast divide-and-conquer sparse Cox regression}.
\bjournal{Biostatistics}
\bvolume{22}
\bpages{381--401}.
\end{barticle}
\endbibitem

\bibitem[\protect\citeauthoryear{Williamson
  et~al.}{2020}]{williamson_factors_2020}
\begin{barticle}[author]
\bauthor{\bsnm{Williamson},~\bfnm{Elizabeth~J.}\binits{E.~J.}},
  \bauthor{\bsnm{Walker},~\bfnm{Alex~J.}\binits{A.~J.}},
  \bauthor{\bsnm{Bhaskaran},~\bfnm{Krishnan}\binits{K.}},
  \bauthor{\bsnm{Bacon},~\bfnm{Seb}\binits{S.}},
  \bauthor{\bsnm{Bates},~\bfnm{Chris}\binits{C.}},
  \bauthor{\bsnm{Morton},~\bfnm{Caroline~E.}\binits{C.~E.}},
  \bauthor{\bsnm{Curtis},~\bfnm{Helen~J.}\binits{H.~J.}},
  \bauthor{\bsnm{Mehrkar},~\bfnm{Amir}\binits{A.}},
  \bauthor{\bsnm{Evans},~\bfnm{David}\binits{D.}},
  \bauthor{\bsnm{Inglesby},~\bfnm{Peter}\binits{P.}},
  \bauthor{\bsnm{Cockburn},~\bfnm{Jonathan}\binits{J.}},
  \bauthor{\bsnm{McDonald},~\bfnm{Helen~I.}\binits{H.~I.}},
  \bauthor{\bsnm{MacKenna},~\bfnm{Brian}\binits{B.}},
  \bauthor{\bsnm{Tomlinson},~\bfnm{Laurie}\binits{L.}},
  \bauthor{\bsnm{Douglas},~\bfnm{Ian~J.}\binits{I.~J.}},
  \bauthor{\bsnm{Rentsch},~\bfnm{Christopher~T.}\binits{C.~T.}},
  \bauthor{\bsnm{Mathur},~\bfnm{Rohini}\binits{R.}},
  \bauthor{\bsnm{Wong},~\bfnm{Angel Y.~S.}\binits{A.~Y.~S.}},
  \bauthor{\bsnm{Grieve},~\bfnm{Richard}\binits{R.}},
  \bauthor{\bsnm{Harrison},~\bfnm{David}\binits{D.}},
  \bauthor{\bsnm{Forbes},~\bfnm{Harriet}\binits{H.}},
  \bauthor{\bsnm{Schultze},~\bfnm{Anna}\binits{A.}},
  \bauthor{\bsnm{Croker},~\bfnm{Richard}\binits{R.}},
  \bauthor{\bsnm{Parry},~\bfnm{John}\binits{J.}},
  \bauthor{\bsnm{Hester},~\bfnm{Frank}\binits{F.}},
  \bauthor{\bsnm{Harper},~\bfnm{Sam}\binits{S.}},
  \bauthor{\bsnm{Perera},~\bfnm{Rafael}\binits{R.}},
  \bauthor{\bsnm{Evans},~\bfnm{Stephen J.~W.}\binits{S.~J.~W.}},
  \bauthor{\bsnm{Smeeth},~\bfnm{Liam}\binits{L.}} \AND
  \bauthor{\bsnm{Goldacre},~\bfnm{Ben}\binits{B.}}
(\byear{2020}).
\btitle{Factors associated with {COVID}-19-related death using {OpenSAFELY}}.
\bjournal{Nature}
\bvolume{584}
\bpages{430--436}.
\bnote{Number: 7821 Publisher: Nature Publishing Group}.
\bdoi{10.1038/s41586-020-2521-4}
\end{barticle}
\endbibitem

\bibitem[\protect\citeauthoryear{Witten and
  Tibshirani}{2010}]{witten_survival_2010}
\begin{barticle}[author]
\bauthor{\bsnm{Witten},~\bfnm{Daniela~M}\binits{D.~M.}} \AND
  \bauthor{\bsnm{Tibshirani},~\bfnm{Robert}\binits{R.}}
(\byear{2010}).
\btitle{Survival analysis with high-dimensional covariates}.
\bjournal{Statistical methods in medical research}
\bvolume{19}
\bpages{29--51}.
\bdoi{10.1177/0962280209105024}
\end{barticle}
\endbibitem

\bibitem[\protect\citeauthoryear{Yang and Zou}{2013}]{yang_cocktail_2013}
\begin{barticle}[author]
\bauthor{\bsnm{Yang},~\bfnm{Yi}\binits{Y.}} \AND
  \bauthor{\bsnm{Zou},~\bfnm{Hui}\binits{H.}}
(\byear{2013}).
\btitle{A cocktail algorithm for solving the elastic net penalized Cox’s
  regression in high dimensions}.
\bjournal{Statistics and its Interface}
\bvolume{6}
\bpages{167--173}.
\end{barticle}
\endbibitem

\bibitem[\protect\citeauthoryear{Yusuf et~al.}{2020}]{yusuf_modifiable_2020}
\begin{barticle}[author]
\bauthor{\bsnm{Yusuf},~\bfnm{Salim}\binits{S.}},
  \bauthor{\bsnm{Joseph},~\bfnm{Philip}\binits{P.}},
  \bauthor{\bsnm{Rangarajan},~\bfnm{Sumathy}\binits{S.}},
  \bauthor{\bsnm{Islam},~\bfnm{Shofiqul}\binits{S.}},
  \bauthor{\bsnm{Mente},~\bfnm{Andrew}\binits{A.}},
  \bauthor{\bsnm{Hystad},~\bfnm{Perry}\binits{P.}},
  \bauthor{\bsnm{Brauer},~\bfnm{Michael}\binits{M.}},
  \bauthor{\bsnm{Kutty},~\bfnm{Vellappillil~Raman}\binits{V.~R.}},
  \bauthor{\bsnm{Gupta},~\bfnm{Rajeev}\binits{R.}},
  \bauthor{\bsnm{Wielgosz},~\bfnm{Andreas}\binits{A.}},
  \bauthor{\bsnm{AlHabib},~\bfnm{Khalid~F}\binits{K.~F.}},
  \bauthor{\bsnm{Dans},~\bfnm{Antonio}\binits{A.}},
  \bauthor{\bsnm{Lopez-Jaramillo},~\bfnm{Patricio}\binits{P.}},
  \bauthor{\bsnm{Avezum},~\bfnm{Alvaro}\binits{A.}},
  \bauthor{\bsnm{Lanas},~\bfnm{Fernando}\binits{F.}},
  \bauthor{\bsnm{Oguz},~\bfnm{Aytekin}\binits{A.}},
  \bauthor{\bsnm{Kruger},~\bfnm{Iolanthe~M}\binits{I.~M.}},
  \bauthor{\bsnm{Diaz},~\bfnm{Rafael}\binits{R.}},
  \bauthor{\bsnm{Yusoff},~\bfnm{Khalid}\binits{K.}},
  \bauthor{\bsnm{Mony},~\bfnm{Prem}\binits{P.}},
  \bauthor{\bsnm{Chifamba},~\bfnm{Jephat}\binits{J.}},
  \bauthor{\bsnm{Yeates},~\bfnm{Karen}\binits{K.}},
  \bauthor{\bsnm{Kelishadi},~\bfnm{Roya}\binits{R.}},
  \bauthor{\bsnm{Yusufali},~\bfnm{Afzalhussein}\binits{A.}},
  \bauthor{\bsnm{Khatib},~\bfnm{Rasha}\binits{R.}},
  \bauthor{\bsnm{Rahman},~\bfnm{Omar}\binits{O.}},
  \bauthor{\bsnm{Zatonska},~\bfnm{Katarzyna}\binits{K.}},
  \bauthor{\bsnm{Iqbal},~\bfnm{Romaina}\binits{R.}},
  \bauthor{\bsnm{Wei},~\bfnm{Li}\binits{L.}},
  \bauthor{\bsnm{Bo},~\bfnm{Hu}\binits{H.}},
  \bauthor{\bsnm{Rosengren},~\bfnm{Annika}\binits{A.}},
  \bauthor{\bsnm{Kaur},~\bfnm{Manmeet}\binits{M.}},
  \bauthor{\bsnm{Mohan},~\bfnm{Viswanathan}\binits{V.}},
  \bauthor{\bsnm{Lear},~\bfnm{Scott~A}\binits{S.~A.}},
  \bauthor{\bsnm{Teo},~\bfnm{Koon~K}\binits{K.~K.}},
  \bauthor{\bsnm{Leong},~\bfnm{Darryl}\binits{D.}},
  \bauthor{\bsnm{O'Donnell},~\bfnm{Martin}\binits{M.}},
  \bauthor{\bsnm{McKee},~\bfnm{Martin}\binits{M.}} \AND
  \bauthor{\bsnm{Dagenais},~\bfnm{Gilles}\binits{G.}}
(\byear{2020}).
\btitle{Modifiable risk factors, cardiovascular disease, and mortality in 155
  722 individuals from 21 high-income, middle-income, and low-income countries
  ({PURE}): a prospective cohort study}.
\bjournal{The Lancet}
\bvolume{395}
\bpages{795--808}.
\bdoi{10.1016/S0140-6736(19)32008-2}
\end{barticle}
\endbibitem

\bibitem[\protect\citeauthoryear{Zhang and Lu}{2007}]{zhang_adaptive_2007}
\begin{barticle}[author]
\bauthor{\bsnm{Zhang},~\bfnm{Hao~Helen}\binits{H.~H.}} \AND
  \bauthor{\bsnm{Lu},~\bfnm{Wenbin}\binits{W.}}
(\byear{2007}).
\btitle{Adaptive {Lasso} for {Cox}'s proportional hazards model}.
\bjournal{Biometrika}
\bvolume{94}
\bpages{691--703}.
\bdoi{10.1093/biomet/asm037}
\end{barticle}
\endbibitem

\bibitem[\protect\citeauthoryear{Zou}{2006}]{zou_adaptive_2006}
\begin{barticle}[author]
\bauthor{\bsnm{Zou},~\bfnm{Hui}\binits{H.}}
(\byear{2006}).
\btitle{The {Adaptive} {Lasso} and {Its} {Oracle} {Properties}}.
\bjournal{Journal of the American Statistical Association}
\bvolume{101}
\bpages{1418--1429}.
\bnote{Publisher: Taylor \& Francis \_eprint:
  https://doi.org/10.1198/016214506000000735}.
\bdoi{10.1198/016214506000000735}
\end{barticle}
\endbibitem

\end{thebibliography}


\end{document}